\let\latexarabic\arabic
\let\latexdocument\document
\let\latexenddocument\enddocument

\RequirePackage[thmmarks]{ntheorem}
\makeatletter
\renewtheoremstyle{plain}
  {\item[\hskip\labelsep \theorem@headerfont ##1\ \textup{##2}\theorem@separator]}
  {\item[\hskip\labelsep \theorem@headerfont ##1\ \textup{##2}\ (##3)\theorem@separator]}
\makeatother

\documentclass[lineno]{biometrika}
 
\usepackage[OT1]{fontenc}

\usepackage[normalem]{ulem} % either use this (simple) or
\usepackage{todonotes}

\let\document\latexdocument
\let\enddocument\latexenddocument
\AtEndDocument{\printhistory}
\let\arabic\latexarabic
\def\rm{}

\usepackage{xcolor}
\usepackage{soul}
\usepackage{cancel}
\usepackage{amsmath}
\usepackage{url}
\usepackage{float}
\usepackage{times}
\usepackage{bm}
\usepackage{natbib}
\usepackage{subcaption}
\usepackage{amssymb}
\usepackage[plain,noend]{algorithm2e}
\usepackage[mathscr]{euscript}
\usepackage{subfiles}

\makeatletter
\renewcommand{\algocf@captiontext}[2]{#1\algocf@typo. \AlCapFnt{}#2} % text of caption
\def\@algocf@capt@plain{top}
\renewcommand{\algocf@makecaption}[2]{%
  \addtolength{\hsize}{\algomargin}%
  \sbox\@tempboxa{\algocf@captiontext{#1}{#2}}%
  \ifdim\wd\@tempboxa >\hsize%     % if caption is longer than a line
    \hskip .5\algomargin%
    \parbox[t]{\hsize}{\algocf@captiontext{#1}{#2}}% then caption is not centered
  \else%
    \global\@minipagefalse%
    \hbox to\hsize{\box\@tempboxa}% else caption is centered
  \fi%
  \addtolength{\hsize}{-\algomargin}%
}
\makeatother

\newcommand{\T}{{\vec{\mathcal{T}}}}

\newcommand{\diag}{\text{diag}}

\newcommand{\bk}{\bar{k}}
\newcommand{\bK}{\mathscr{K}}

\protected\def\[#1\]{\begin{equation}\begin{aligned}#1\end{aligned}\end{equation}}
\protected\def\(#1\){\begin{equation*}\begin{aligned}#1\end{aligned}\end{equation*}}

\usepackage{overpic}

\begin{document}

\jname{Manuscript}
% \jname{Manuscript}
%% The year, volume, and number are determined on publication

% \jyear{2024}
% \jvol{103}
% \jnum{1}

%% The \doi{...} and \accessdate commands are used by the production team
%\doi{10.1093/biomet/asm023}

% \accessdate{Advance Access publication on 31 July 2016}

%% These dates are usually set by the production team
% \received{2 January 2017}
% \revised{1 April 2017}

%% The left and right page headers are defined here:
\markboth{E. Tam, D.B. Dunson and L.L. Duan}{Exact Sampling of Spanning Trees via Fast-forwarded Random Walks}

%% Here are the title, author names and addresses
\title{Exact Sampling of Spanning Trees\\ via Fast-Forwarded Random Walks}

\author{Edric Tam}
\affil{Department of Biomedical Data Science, Stanford University, 
300 Pasteur Drive, Palo Alto, California 94304, U.S.A. \email{edrictam@stanford.edu}}

\author{David B. Dunson}
\affil{Department of Statistical Science and Department of Mathematics, Duke University, Box 90251
Durham, North Carolina 27708, U.S.A. \email{dunson@duke.edu}}

\author{\and Leo L. Duan}
\affil{Department of Statistics, University of Florida, 101C Griffin-Floyd Hall, P.O. Box 118545, Gainesville, Florida 32611, U.S.A.\email{li.duan@ufl.edu}}

\maketitle

\begin{abstract}

Tree graphs are used routinely in statistics. When estimating a Bayesian model with a tree component, sampling the posterior remains a core difficulty. Existing Markov chain Monte Carlo methods tend to rely on local moves, often leading to poor mixing. A promising approach is to instead directly sample spanning trees on an auxiliary graph. Current spanning tree samplers, such as the celebrated Aldous--Broder algorithm, rely predominantly on simulating random walks that are required to visit all the nodes of the graph. Such algorithms are prone to getting stuck in certain sub-graphs. We formalize this phenomenon using the bottlenecks in the random walk's transition probability matrix. We then propose a novel fast-forwarded cover algorithm that can break free from bottlenecks. The core idea is a marginalization argument that leads to a closed-form expression which allows for fast-forwarding to the event of visiting a new node. Unlike many existing approximation algorithms, our algorithm yields exact samples. We demonstrate the enhanced efficiency of the fast-forwarded cover algorithm, and illustrate its application in fitting a Bayesian dendrogram model on a Massachusetts crimes and communities dataset. 
\end{abstract}

\begin{keywords}
Aldous--Broder algorithm;
Bayesian modeling;
Isoperimetric constant; Random walk; Invariant distribution; Spectral graph theory.
\end{keywords}

\section{Introduction}
Tree graphs are commonly encountered in statistical modeling. An undirected tree is an acyclic and connected graph $T=(V_T,E_T)$ with nodes $V_T=(1,\ldots, m)$ and edges $E_T=\{(j,l)\}$ with $|E_T|=m-1$. By designating a node $r \in V_T$ as root, one can easily obtain from $T$ a directed tree $\vec T = (V_{\vec T},E_{\vec T}, r)$, with an identical node set $V_{\vec T} = V_T$ and directed edges $E_{\vec T}=\{(j\to l)\}$ obtained from $E_T$ by pointing the edges away from $r$. Such tree structures provide succinct ways to capture complex dependencies that arise from a wide range of statistical applications. 

In hierarchical modeling, directed trees represent multi-layer dependence structures underlying the observed data. From a generative perspective, we consider a $\vec T$ 
where each node $v$ is equipped with a parameter $\mu_v$. Using this tree, we define an augmented likelihood for data $y_1,\ldots, y_n$ and parameters $\mu_1,\ldots,\mu_m$ conditioned on assignment labels $z_i\in (1,\ldots, m)$:

\begin{eqnarray}
L(y,\mu\mid  \vec T, z)  = \Big \{ \prod_{i=1}^n\mathcal F (y_i \mid \mu_{z_i} ) \Big\} \Big \{ \mathcal R(\mu_r) \prod_{(j\to l)\in E_{\vec { T}}} \mathcal H(\mu_l\mid \mu_j) \Big \},
\label{eq:BayesDendro}
\end{eqnarray}
where $\mathcal H(\mu_l\mid \mu_j)$ is the transition probability kernel from $\mu_j$ to $\mu_l$, $\mathcal R(\mu_r)$ is the marginal kernel for $\mu_r$ in the root, and $\mathcal F$ is the conditional kernel of the data $y_i$ given the assignment $z_i$ for that observation. The $\vec T$ that we condition on includes both the edge set $E_{\vec T}$ and the root node $r$. There are potentially other parameters characterizing $\mathcal F$ and $\mathcal H$, including dependence on covariates via decision trees \citep{chipman1998bayesian,castillo2021uncertainty} or related ensemble methods \citep{chipman2010bart,linero2018bayesian}, but we suppress these temporarily for ease of notation. Model \eqref{eq:BayesDendro} induces a partition on $(1,\ldots,n)$ via the latent assignments $z$. In contrast to traditional mixture models, which often assume the $\mu_k$'s to be generated independently from a common distribution, \eqref{eq:BayesDendro} characterizes the dependence in $\mu_j$ and $\mu_l$ through an ancestry tree. Ancestors of $v$ include the tree nodes in the path from the root to $v$. This type of dependence is well motivated in many application areas. The tree can be interpreted as an inferred evolutionary/phylogenetic history in certain biological settings \citep{huelsenbeck2001mrbayes,suchard2001bayesian,neal2003density}, or alternatively as a multi-layer partitioning of a dataset \citep{heller2005bayesian}.

It is also common to use a collection, or forest, of trees for flexibility in modeling. Consider 
\begin{eqnarray}
L(y\mid  \vec T_1, \ldots, \vec T_{\bK}) = \prod_{\bk=1}^{\bK} \bigg \{\mathcal R(y_{r^{(\bk)}}) \prod_{(j \to  l)\in E_{\vec T_{\bk}}} \mathcal H(y_l\mid y_j) \bigg \},
\label{eq:forest}
\end{eqnarray}
where each $\vec T_{\bk}= \{ V_{\vec T_{\bk}}, E_{\vec T_{\bk}}, r^{(\bk)} \}$ is a component tree, and $(V_{\vec T_1},\ldots, V_{\vec T_{\bK}})$ gives a $\bK$-partition of data index $(1,\ldots, n)$, where $n = \sum_{{\bk} = 1}^{\bK} |V_{\vec T_{\bk}}|$. The kernel $\mathcal R$ describes how the first point in a group arises, and $\mathcal H$ characterizes the conditional dependence of the subsequent points given the previous ones. 
The use of trees and forests in graphical modeling  \citep{lauritzen2011elements} dates back at least to the single-linkage clustering algorithm \citep{gower1969minimum}, and is recently seen in dependence graph estimation \citep{duan2021bayesian}, contiguous spatial partitioning \citep{teixeira2019bayesian,luo2021bayesian,luo2023nonstationary}, and model-based spectral clustering \citep{duan2023spectral}. In addition, likelihood \eqref{eq:forest} has been extended to a mixture of overlapping trees in Bayesian network estimation  \citep{meila2000learning,meilua2006tractable,elidan2008learning}.

Although there is a rich literature on algorithmic approaches for obtaining point estimates of tree graphs \citep{prim1957shortest,kruskal1956shortest}, we are particularly interested in model-based Bayesian approaches. Such methods have the advantage of providing a characterization of uncertainty in estimating trees, while also inferring a generative probability model for the data. Quantification of uncertainty is crucial in this context. Algorithms that produce a single tree estimate are ripe for over-interpretation and lack of reproducibility, since in most applications there are many different trees that are almost equally plausible for the data. Naturally, the success of such a Bayesian approach hinges on whether one can
conduct inferences based on the posterior distribution of trees in a computationally efficient way.

Sampling from posterior distributions for trees is generally a difficult problem. Current Markov chain Monte Carlo samplers often navigate tree spaces using local modifications, such as pruning and growing moves. Since tree spaces are combinatorial and large in size, these samplers often exhibit poor mixing. A natural alternative is to rely on conjugacy to employ block updates on $\vec T$ in Gibbs-type samplers. Our strategy
is to view $\vec T$ as a {\em spanning tree} $\vec{\mathcal T}$ under a complete and weighted auxiliary graph $G=(V_G,E_G)$.  Here, a spanning tree $\vec{\mathcal T}$ of $G$ is simply a subtree of $G$ that spans all the nodes of $G$ with edges oriented away from some root node $r$. For the generative process in \eqref{eq:BayesDendro}, one can consider a complete graph $G$ with nodes $(1,\ldots, m)$, with a weighted adjacency matrix $Q\in[0,\infty)^{m \times m }$ of elements  $q_{j, l} = \mathcal H(\mu_l\mid \mu_j)$ 
 for every pair $(j,l)$. Observe that equation \eqref{eq:BayesDendro} can be rewritten as $L(y,\mu \mid  \vec{\mathcal{T}}, z)  =   \mathcal R(\mu_r)\prod_{(j\to l)\in E_{\vec{\mathcal{T}}}} \Big \{  \prod_{z_i= l}\mathcal F (y_i \mid \mu_{z_i} ) \mathcal H(\mu_l\mid \mu_j) \Big \}$. Under a uniform prior $\Pi_0(\vec{\mathcal{T}}) \propto 1$,  we can conduct a full conditional update from $\Pi(\vec {\mathcal T}\mid -)$ by drawing from the following two  distributions
 
\[
\label{eq:product_prob}
\text{Pr}(r) = \frac{g_r/\rho_r}{\sum_{r'=1}^{m}g_{r'}/\rho_{r'}}, \qquad
\text{Pr}(\vec {\mathcal T} | r) = \rho_r \prod_{(j\to l)\in E_{ \vec{\mathcal T}}} q_{j,l},
\]
Here, $\text{Pr}(r)$ is a discrete probability distribution on $(1,\ldots,m)$ and $\text{Pr}(\vec {\mathcal T} | r)$ is a distribution over the edges of $\vec {\mathcal T}$ given a root node $r$. We define $g_r=\mathcal R(\mu_r)$. The term
$\rho_r = \{\sum_{\vec{\mathcal T}\text{ rooted at }r}\prod_{(j\to l)\in E_{ \vec{\mathcal T}}} q_{j,l}\}^{-1}$ is a normalization constant in $\text{Pr}(\vec {\mathcal T} | r)$ that could vary with $r$. The uniform prior $\Pi_0(\vec{\mathcal{T}}) \propto 1$ is a special case of a more general class of conjugate priors, consisting of a root term multiplied by a product on the edges of the spanning tree. We discuss the sampling of $\text{Pr}(r)$ under various scenarios in Section 2.4.
Here, the imperative is to efficiently sample from spanning tree distributions of the form $\text{Pr}(\vec {\mathcal T} | r)$.

\begin{figure}[H]
    \centering
        \includegraphics[width=0.5\textwidth]{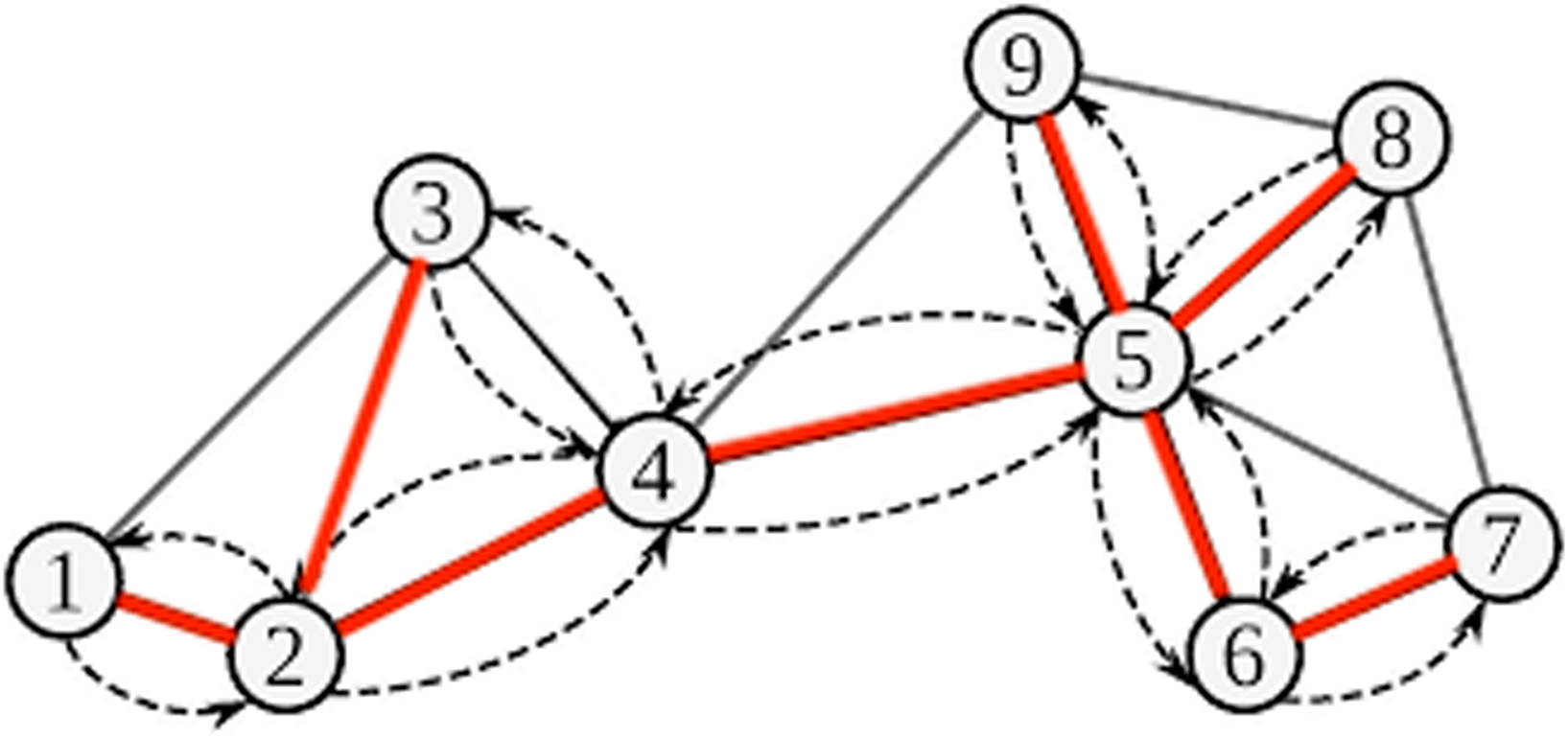}
\caption{Illustration for sampling a spanning tree. The edges of the underlying connected graph are shown in solid black lines. We start from a root node $r$ (drawn proportional to $g_r/\rho_r$) and perform a random walk (dashed arrows) until we visit all nodes. The nodes are labeled by the order in which they are visited. The first entrance edges (red lines) of each node together form a spanning tree $\vec{\mathcal T}$.
}
\label{fig:spanning_illustration}
\end{figure}

The celebrated Aldous--Broder algorithm \citep{aldous1990random,broder1989generating} provides a tractable way to draw exact samples from $\text{Pr}(\vec {\mathcal T} | r) $. Figure \ref{fig:spanning_illustration} provides an illustration.
The algorithm proceeds by taking a random walk $X_t=(r,x_1,x_2,\ldots, x_t)$ on $V_G$ and stops at the cover time of the walk when all nodes have been visited. By collecting the first entrance edges of $X_t$, which are the set of edges via which the walk first visited each node, one obtains a spanning tree $\vec{\mathcal{T}}$ rooted at $r$. By separately specifying a distribution on the root node $r$, a distribution is obtained on the directed spanning trees $\vec {\mathcal T}$.
 For a graph $G$ containing $m$ nodes, \cite{broder1989bounds} shows an expected cover time of $O(m\log m)$ in well-connected graphs and $O(m^3)$ in the worst case. 

The Aldous--Broder algorithm can be empirically slow. In practice, even if the number of nodes $m$ in $G$ is small, the algorithm can get stuck in certain subgraphs for a long time. This issue is inherent to the nature of random walks. Other popular algorithms, such as Wilson's algorithm based on the loop-erased random walk \citep{wilson1996generating}, also suffer from the same problem. Beyond random walks, there are Laplacian-based algorithms that sample uniform spanning trees from unweighted graphs \citep{guenoche1983random, kulkarni1990Generating, harvey2016Generating}. There are also approximate random spanning tree samplers \citep{madry2014fast,schild2018almost} that rely on Laplacian solvers. In this article, we focus on algorithms based on random walks for their more general applicability.

We formalize the pathological phenomenon of the Aldous--Broder algorithm using eigenvalues of a normalized Laplacian to characterize bottlenecks in the graph. We then propose a fast-forwarded cover algorithm for exact sampling from $\text{Pr}(\vec {\mathcal T} | r) $ while bypassing wasteful random walk steps. The key observation is that we do not need to simulate the entire random walk trajectory to obtain the first entrance edges to each node. We derive a closed form expression that allows for direct, fast-forwarded sampling of these first entrance edges via a marginalization argument. In simulations, our fast-forwarded sampler enjoys much faster empirical performance against competitors when bottlenecks are present. 

Existing spanning tree samplers often perform transitions directly according to the underlying graph's adjacency matrix $Q$, restricting their applicability to symmetric/undirected cases. We show that by using an auxiliary matrix $W$, which is generated from $Q$ under certain transformations, to specify the random walk transition probabilities, our fast-forwarded cover algorithm can be extended to more general scenarios, including cases where the underlying graph has an asymmetric adjacency matrix or the induced Markov chains are irreversible. These results are of independent interest. We illustrate our sampler by fitting a Bayesian dendrogram model on a Massachusetts crime and community dataset. The resulting Gibbs sampler exhibits drastically improved mixing performance compared to a reversible jump sampler based on local moves and a sampler based on subtree prune and regraft moves. 
\section{Method}
\subsection{Background on Aldous--Broder Algorithm}
We review and summarize results on the Aldous--Broder algorithm \citep{aldous1990random,broder1989generating} in this section. This algorithm samples a random spanning tree $\vec{\mathcal T}$ from the underlying graph $G$ based on a random walk. Consider a weight matrix $W\in [0,\infty)^{m\times m}$ that is used to specify the probabilities of random walk transitions on $G$. For simplicity, it suffices for now to consider $W$ as the graph's weighted adjacency matrix $Q$ in \eqref{eq:product_prob}, an important special case that applies when $Q$ is symmetric. However, our results hold for more general cases where $W$ is some specific transformation of $Q$. A detailed discussion is given in Section 2.4.

We use $W$ to specify a random walk over the state space $V_G=(1,\ldots,m)$. This random walk is a discrete-time Markov chain $X=(x_0,x_1,\ldots)$ with transition probabilities:
\[
p_{j,l}=\textup{Pr}(x_{t+1}=l \mid x_{t}=j)= \frac{w_{j,l}}{d_j}, \qquad d_j=\sum_{v=1}^m w_{j,v},
\]
where $d_j>0$ and $w$ denotes entries of $W$. We assume that the Markov chain is irreducible: for any pair of $(j,l)$, there exists $t>0$ such that $\textup{Pr}(x_{t}=l \mid x_{0}=j)>0$. We denote the random walk up to time $t$ by $X_t=(x_0,x_1,\ldots, x_t)$, and the invariant distribution of $x_t:t\to \infty$ by $(\pi_1,\ldots, \pi_m)$. The result of \cite{broder1989generating} further assumes reversibility of the Markov chain, $\pi_j p_{j,l} = \pi_l p_{l,j}$; however, this condition is not needed here.  
 
The Aldous--Broder algorithm proceeds as follows: initialize the walk $x_0$ at root $r$; perform the random walk until all nodes are visited at the cover time $\hat t$; construct a directed spanning tree $\vec{\mathcal T}$ with the edge set as the set of first-entrance edges all pointed away from $r$. Formally, 
\[\label{eq:first_ent}
E_{\vec {\mathcal T}} = \bigcup_{j\in (V_G\setminus r)} \Big\{ (x_{ (t_j-1)}\to x_{t_j}):t_j=\min_{1\le t\le \hat t} (t: x_t=j)\Big\}.
\]
The following theorem characterizes the induced distribution of $\T$.
\begin{theorem}[Extended Aldous--Broder]\label{aldousbroderthm} Letting $\T$ be generated as above, then
\[\label{eq:ext_ab}
\textup{Pr}( \T \mid r) 
\propto \Big\{ \prod_{(j\to l) \in E_{\vec {\mathcal T}}}\frac{p_{j,l}\pi_j}{\pi_l} \Big\} 
\frac{1}{\pi_r} 
\propto \prod_{(j\to l) \in E_{\vec {\mathcal T}}}({p_{j,l}\pi_j} ),
\]
where the probability is normalized over all directed spanning trees $\T$ rooted from $r$.
\end{theorem} If the underlying graph $G$ is directed, the choice of root $r$ affects the transition probabilities involved in sampling $\vec {\mathcal T}$. We discuss details on sampling $r$ in Section 2.4.
The first term in \eqref{eq:ext_ab} is based on an adaptation of Theorem 3 of \cite{fredes2023combinatorial}, where a directed tree with edges pointed towards the root is used, as is standard in the probability literature.  While this is the opposite of our choice of orientation, the mathematical results carry over. The second term is based on the fact that $\{\prod_{(j\to l) \in E_{\vec {\mathcal T}}}{\pi_l} \} \pi_r= \prod_{j=1}^m \pi_j$, which is invariant to $E_{\T}$ and hence can be omitted. 

In Section 2.4, we will also describe how one can choose $W$ as some appropriate transformation of $Q$ so that \eqref{eq:ext_ab}  becomes the product form shown in \eqref{eq:product_prob} in general settings without requiring matrix symmetry for $W$. For now in Section 2.2, we focus on computational aspects and discuss bottleneck effects that impact the cover time of the Aldous--Broder algorithm.

\subsection{Bottlenecks in Random Walk Cover Algorithm}

The Aldous--Broder algorithm is only efficient if there is a short cover time $\hat t$ to reach all nodes in $V_G$. However, when there are bottlenecks in the graph that lead to close to zero probabilities to move from the visited nodes $U$ to the unvisited nodes $\bar U=V_G\setminus U$, the random walk can spend a long time within $U$.  We now characterize the consequences of this curse-of-bottlenecks. 

\begin{figure}[H]
    \centering
    \begin{subfigure}[t]{0.32\textwidth}
        \centering
        \includegraphics[width=1\textwidth]{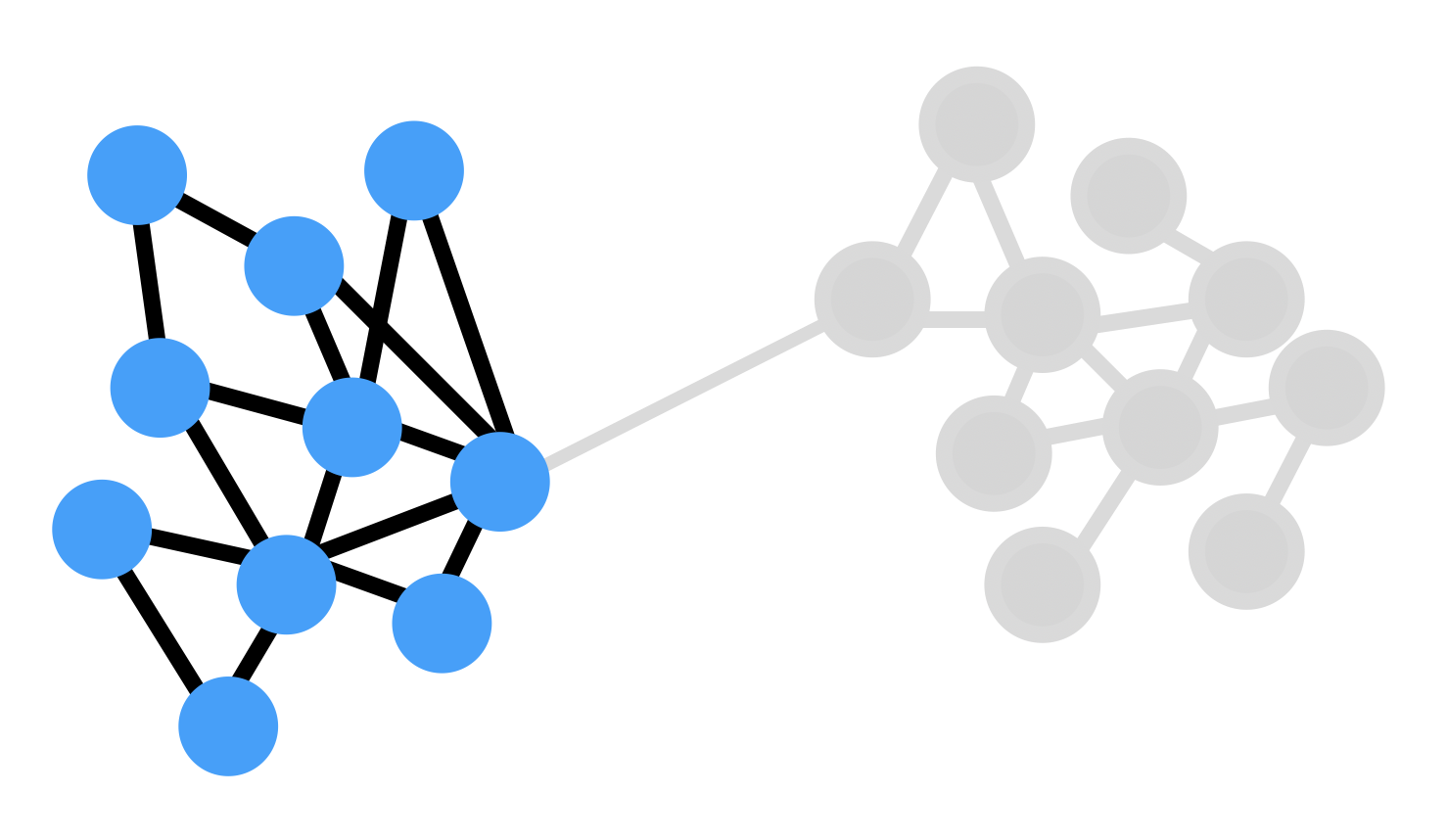}
        \caption{Low edge weight between  visited nodes (blue) and unvisited (grey).}
    \end{subfigure}\;
    \begin{subfigure}[t]{0.32\textwidth}
        \centering
        \includegraphics[width=0.7\textwidth]{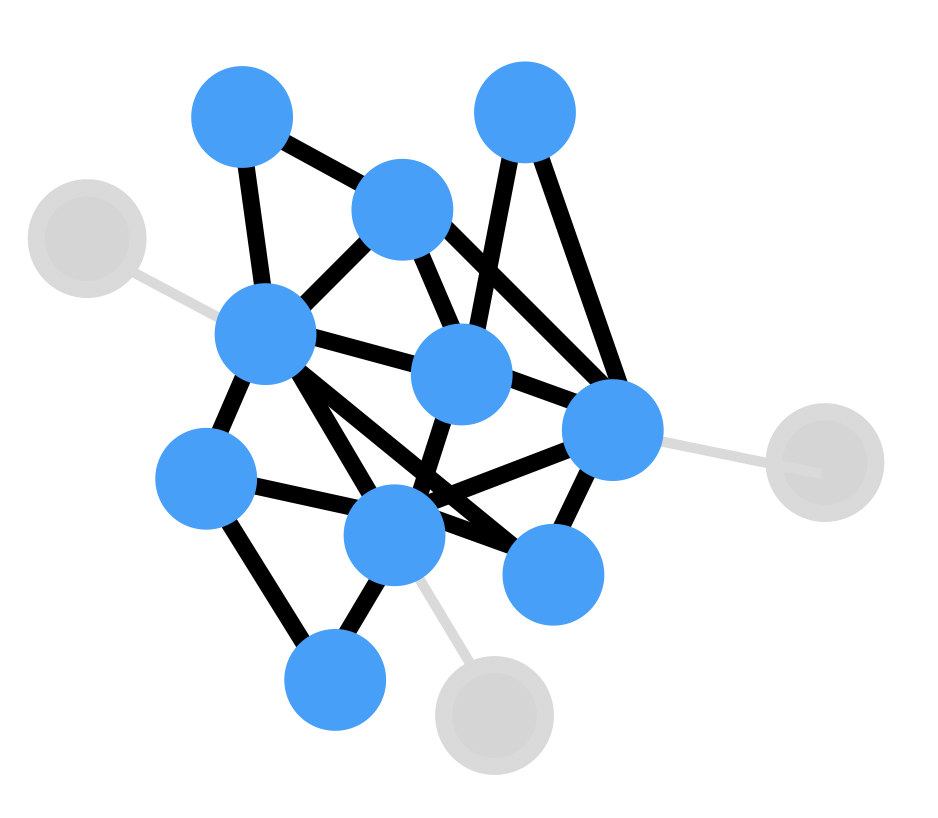}
        \caption{Unvisited nodes (grey) that are isolated.}
    \end{subfigure}\;
    \begin{subfigure}[t]{0.32\textwidth}
        \centering
        \includegraphics[width=.9\textwidth]{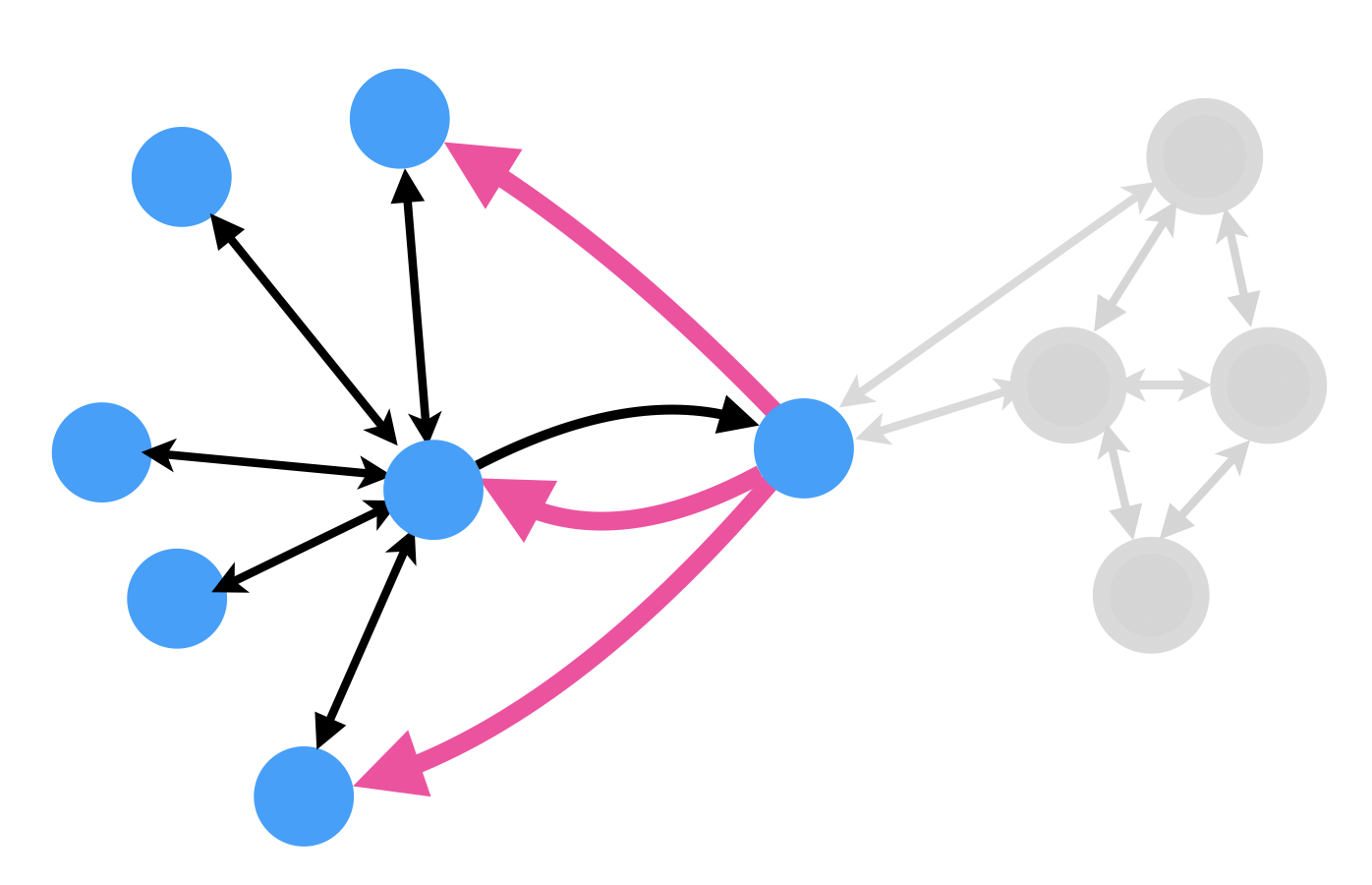}
        \caption{Large probability (magenta lines) to return to visited nodes (blue) in a directed graph.}
    \end{subfigure}
    \caption{Illustration of some commonly encountered graphs with curse of bottlenecks --- a low transition probability to move from the visited nodes (blue) to those unvisited (grey).}
    \label{fig:curse}
\end{figure}

Figure \ref{fig:curse} illustrates three common challenges through example graphs. The first two graphs are either directed or undirected, while the third is directed. First, as in panel (a), there may be disjoint sets of nodes that are only weakly connected, in the sense that the transition probability between these sets is small. Second, as in panel (b), there may be nodes that are isolated with few edges incident to them, and each edge having low transition probability. In this case, the random walk cover algorithm may visit most of the nodes within a short time, but faces difficulties reaching the last few. Third, as in panel (c), in a directed graph, due to edge weight asymmetry, there may be a much higher probability to remain in a node set that has been visited. 

Regardless of the specific scenario, the curse-of-bottlenecks phenomenon can be understood as a close-to-zero probability to move to the unvisited set, marginalized over both the current node and the potential arrival node amongst unvisited nodes. 

We now formally characterize this phenomenon. Consider a strictly increasing node set $(r)=U_1\subset U_2\subset \ldots\subset U_m=V_G$, recording the history of nodes visited by the random walk, with 
$U_k= \cup_{t=0}^{\hat t_k}  X_t$ and 
$\hat t_k= \min ( t\ge 0:|\cup_{\tilde t=0}^t  X_{\tilde t}|=k),$
where we refer to $\hat t_k$  as the partial cover time for $k$ nodes. 
The probability of transitioning outside of $U_k$ at time $t+1$, conditional on the walk remaining within $U_k$ from the beginning through time $t$ is 
\(
\text{Pr}(x_{t+1}\in \bar U_k \mid x_{\tilde t\le t} \in U_k) & = 
\sum_{j\in U_k} \bigg \{ \frac{\sum_{l\in \bar U_k} w_{j,l}}{d_j} \text{Pr}(x_t=j \mid x_{\tilde t\le t} \in U_k) \bigg\}\le  
\max_{j\in U_k} \bigg( \frac{\sum_{l\in \bar U_k} w_{j,l}}{
d_j
} \bigg)\\
& = 
\bigg( { \min_{j\in U_k} 
\frac {\sum_{l'\in  U_k} w_{j,l'} } {\sum_{l\in \bar U_k} w_{j,l}}+1
} \bigg)^{-1},
\)
where the inequality uses the fact that the weighted average over $j\in U_k$ is less than or equal to the maximum, and the last equality is based on decomposing $d_j=\sum_{l\in \bar U_k} w_{j,l} +\sum_{l'\in  U_k} w_{j,l'} $. 
Therefore, if the total outgoing weights to $\bar U_k$ are dominated by the weights to stay within $U_k$, in the sense that ${\sum_{l\in \bar U_k} w_{j,l}} \ll  {\sum_{l'\in  U_k} w_{j,l'}}$, it is unlikely that $x_{t+1}\in \bar U_k$. With the above ingredients, we are ready to quantify the expected cover time.
\begin{theorem}
        For a history of visited nodes $(U_1,\ldots,U_m)$, the expected cover time $\hat t=\hat t_m$ has
\(
\mathbb{E} (\hat t \mid U_1,\ldots,U_m)\ge  (m-1)+\sum_{k=1}^{m-1} \min_{j\in U_k}\frac {\sum_{l'\in  U_k} w_{j,l'} } {\sum_{l\in \bar U_k} w_{j,l}} .
\)
\end{theorem}
\begin{remark}\label{rmk:cover_time}
The lower bound applies to any $W\in [0,\infty)^{m\times m}$ that leads to an irreducible Markov chain. In practice, when the random walk cover algorithm appears to be stuck at a certain $U_k$, one can directly use $\min_{j\in U_k}({\sum_{l'\in  U_k} w_{j,l'} })/ ({\sum_{l\in \bar U_k} w_{j,l}})$ as an estimate of the expected time to leave.
\end{remark}

The above result is a quantification for a specific node history of $(U_1,\ldots,U_m)$. One can obtain a worst case expected cover time by maximizing over all possible sequences of $(U_1,\ldots,U_m)$.  
We are mainly interested in $W$ matrices that satisfy a {\em circulation condition}:
$\sum_{l=1}^m w_{j,l}= \sum_{k=1}^m w_{k,j}, j=1,\ldots,m$. This condition can be applied to the directed graph setting, and includes symmetric $W$ with $w_{j,l}=w_{l,j}$ as a special case.

\begin{theorem}\label{th:bottleneck}
For a random walk $X$ based on weight matrix $W$ satisfying the circulation condition, the worst case expected cover time $\hat t$ has
\(
\max_{U_*\subset V_G:\; r\in U_*}
%\min_{\text{all }(U_1,\ldots, U_m)\ni U^*} 
\mathbb{E} \{ \hat t \mid  (U_1,\ldots,U_m)
, U_*\in (U_1,\ldots,U_m)
\}\ge  \frac{1}{M\sqrt{\lambda_2(\mathcal
L)} } + m-2,
\)
with ,
$M=\sup_{t\ge 0}\sup_j \{\text{Pr}(x_t=j)/\pi_j\}$, and $\lambda_2(\mathcal
L)$ the second smallest eigenvalue of the normalized graph Laplacian $\mathcal L$.
\end{theorem}

\begin{remark}
The above worst-case scenario corresponds to
when the sampler visits a certain node subset $\hat U_*\subset V_G$ and gets stuck.
 Although the above lower bound on expected cover time is inspired by earlier works \citep{matthews1988covering,lovasz1993note,levin2017markov}, our result is distinguished by a directly computable lower bound. The constant $M^{-1}$ has a lower bound $\min_j\pi_j$ for any root distribution on $x_0=r$. In the special case where the root node is drawn from the invariant distribution $r\sim\pi$,  $M=1$ since $\text{Pr}(x_t=j)=\pi_j$ for any $t\geq 0$.
\end{remark}

\subsection{Skipping Bottlenecks via Fast-forwarding}

With the curse of bottlenecks on the cover time of random walks established, we now exploit a useful fact ---
when transforming the random walk trajectory $X_{\hat t}$ into the spanning tree edges $E_{\T}$, one only needs the first entrance edges to each node. It is unnecessary to simulate entire random walk trajectories until reaching a new node if the marginal distribution of the next first entrance edge can be directly sampled. We formalize this idea below.

We set up the notation here. At any time $t$, suppose that we have visited the nodes in $U$, and let $e_j^t=1$ if the walk is at node $j \in U$ at time $t$ and the first exits $U$ at time $t + 1$.  Let $e_j^t=0$ otherwise.  
Intuitively, $e^t_j$ represents the decision made at node $j$ and time $t$, on whether to leave $U$ at the next time point $t+1$. These decision probabilities can differ with $j$.

Given a starting location $x_{t_0}=j_0 \in U$ at some time $t_0$, drawing the first entrance edge into $\overline{U}$ at $t_0+\delta$ for some fixed $\delta\in \mathbb{Z}_+$ can be understood as the result of the following events,
\(
( e^{t_0}_{j_0}=0 ) \Rightarrow (x_{{t_0}+1}=j_{1} \in U) \Rightarrow (e^{{t_0}+1}_{j_{1}}=0) \Rightarrow \cdots \Rightarrow (x_{{t_0}+\delta}=j'\in U) \Rightarrow (e^{t_0+\delta}_{j'}=1),
\) 
where $A \Rightarrow B$ represents the event $B$ occurring after $A$.
The trajectory of the walk stays within $U$ from time $t_0$ until $t_0 + \delta$ and exits from node $j'\in U$ to node $l' \in \bar U$ at time $t_0 + \delta + 1$.
This implies the first-entrance edge $j'\to l'$. We write 
\(
&\eta_j = \text{Pr}(e^t_j=1 \mid x_t=j), \qquad (1-\eta_j) = \text{Pr}(e^t_j=0 \mid x_t=j)= \frac{\sum_{k\in U} w_{j,k}}{d_j}. 
\)
Collect the $\eta$'s into vector form $\eta_U=(\eta_j:j\in U)$. 
Let $P_{U,U}= \{ p_{j,l}:j\in U,l\in U\}$ denote the submatrix of transition probabilities from nodes in $U$ back into $U$. 
Writing a state vector $s_{t}=\{ \text{Pr}(x_{t}=j' \mid x_{t_0}=j_{0},
 x_{\tilde{t}}\in U, \tilde{t} < t)\}_{j'\in U}$, we have the recursive relation
$
s_{t+1} = P_{U,U}^{\rm T} s_t,
$
when the walk has not left $U$ by time $t+1$. The initial state $s_{t_0}$ is a $|U|$-dimensional vector with the element corresponding to $j_0$ set to $1$ and all others equal to $0$. 
For notational ease, we omit  $x_{\tilde{t}}\in U, \tilde{t} < t_0 + \delta$ in the conditioning below.
 We then have the vector
\(
& \{\text{Pr}(x_{t_0+\delta}=j', x_{t_0+\delta+1} \in \bar U  \mid x_{t_0}=j_{0})\}_{j'\in U} 
= \diag(\eta_U)  (P_{U,U}^{\rm T} )^{\delta-1} s_{t_0}.\)
Here, we started at the initial state at time $t_0$, transitioned within the visited nodes $U$ for $\delta - 1$ times, and visited $\bar U$ at time $t+\delta + 1$ via $\diag(\eta_U)$. This expression yields a distribution on the nodes $j'$ that $x_{t_0+\delta}$ can take.  We can further marginalize the above by summing over $\delta\in \mathbb{Z}_{\ge 0}$.  Since the summation involves a Neumann series, it has a closed-form for the marginal if the series converges. The following theorem shows a sufficient condition.
\begin{theorem}
If $P_{U,U}$ is irreducible (cannot be rearranged to a block upper triangular matrix by row and column permutations), and there exists at least one $j\in U:\eta_j>0$, then
\[\label{eq:exit}
\{\textup{Pr}(x_{({{\hat t_{|U| + 1}}-1)}}= j'\mid x_{t_0}=j_{0})\}_{j'\in U} =\textup{diag}(\eta_U)  (I-P_{U,U}^{\rm T} )^{-1} s_{t_0},
\]
where $({{\hat t_{|U| + 1}}-1})$ is the time point before moving to a node in $\bar U$.
\end{theorem}

The above irreducibility condition is satisfied when $w_{j, l} > 0$ for all $j \neq l$. This theorem allows us to sample the first exit edge $(j', l')$ from $U$ to $\bar U$ in a straightforward manner. We first draw $j' \in U$ according to the distribution specified by the vector in \eqref{eq:exit}. We then take a random step from the drawn node $j'$ via the transition specified by
\[\label{eq:exit_edge}\text{Pr}(  x_{{{\hat t_{|U| + 1}}}} = l' \mid x_{({{\hat t_{|U| + 1}}-1)}}=j',e^{(\hat t_{|U| + 1}-1)}_j=1)= { w_{j',l'}}/{\sum_{k\in \bar U} w_{j',k}}\] to obtain $l' \in \bar U$. We refer to steps \eqref{eq:exit} and \eqref{eq:exit_edge} as the {\em fast-forwarding steps}.

\begin{remark}
The direct calculation of the matrix inverse  $(I-P_{U,U}^{\rm T})^{-1}$ 
has $O(|U|^{3})$ cost, but solving the linear equation $x:(I-P_{U,U}^{\rm T} )x=s_{t_0}$ using iterative numerical methods can lead to a lower cost $O(\tilde K|U|)$, with $\tilde K$ the number of iterations until convergence \citep{saad2003iterative}.

\end{remark}

A natural algorithm for drawing a spanning tree performs Aldous--Broder type random walk steps until a bottleneck is reached, and then uses 
fast-forwarding steps; this is then repeated until visiting all nodes. We call this approach the \emph{fast-forwarded cover algorithm}.

\begin{algo}
The fast-forwarded cover algorithm.
\vspace*{-6pt}
\begin{tabbing}
   \qquad \enspace Initialize $x_0=r$. Initialize the first entrance step tracker $\alpha = 1$. Initialize $\tau = 1$. \\
   \qquad 1.\enspace  Simulate one random walk step to $x_\tau$\\
   \qquad 2.\enspace If $x_{\tau}$ has not not been visited before, update $\alpha\leftarrow \tau$.\\
   \qquad 3.\enspace If the steps taken since first entrance step $(\tau- \alpha) \ge  \kappa_0$, a pre-set threshold, \\
   \qquad \qquad \enspace update $x_{\tau+1}$ to $j'$ sampled according to \eqref{eq:exit}\\
    \qquad \qquad \enspace update $x_{\tau+2}$ to $l'$ sampled according to  \eqref{eq:exit_edge}\\
    \qquad \qquad \enspace update $\alpha \leftarrow \tau+2$, and $\tau\leftarrow \tau+3$.  Go  to Step 1. \\\
  \qquad \enspace \enspace If $(\tau- \alpha) <  \kappa_0$, update $\tau\leftarrow \tau+1$. Go  to Step 1.\\
  \qquad \enspace Repeat the above until all $\alpha = m$, where $m$ is the number of nodes  of the underlying graph.\\
   \qquad \enspace Collect and return the first entrance edges.\\
\end{tabbing}
\end{algo}

We use a new index $\tau$ above, since the iteration number will be different from the underlying random walk time index $t$, as soon as a fast-forwarding step is used. The algorithm completes when all nodes have been visited, and the number of iterations required is at most $ \kappa_0(m-1)$. 
In this article, we set $\kappa_0=1,000$, which leads to excellent performance in all our experiments. More generally, since the fast-forwarding step has complexity $O(\tilde K|U|)$, it is natural to consider a varying threshold $\kappa_\tau \propto m_\tau$, where $m_\tau$ is the number of visited nodes $|U|$ by iteration $\tau$.

Before presenting our empirical results, we generalize the random walk cover and our fast-forwarded modifications to be broadly applicable to sampling any directed spanning tree with probability in the form of \eqref{eq:product_prob}. This is accomplished through a careful specification of the transition matrix $P$, or equivalently $W$ up to row-wise normalization. 

\subsection{Sampling from Root Distribution and Applicability of Cover Algorithms}

Given a graph with weighted adjacency matrix $Q$, our goal is to draw samples from \eqref{eq:product_prob} using cover algorithms, a terminology we use to include random walk cover and our fast-forwarded modifications.
We now discuss discuss how to sample $\text{Pr}(r) \propto  g_r/\rho_r$.
Two cases are of interest. 

\noindent\textbf{Case 1, Circulation:} We start with the canonical case in which a simple condition $\sum_{l=1}^m q_{j,l}= \sum_{k=1}^m q_{k,j}$ holds for $j=1,\ldots,m$. If we view $q_{j,l}$ as a flow from node $j$ to $l$, these equalities describe a type of flow conservation, with $Q$ describing a circulation matrix \citep{chung2005laplacians}. This includes the special case where $Q$ is symmetric. Under this circulation condition, setting the random walk transition probabilities  $w_{j,l}$ as equal to the underlying graph's edge weights $q_{j,l}$ is sufficient for cover algorithms to produce samples satisfying
\[\label{eq:global_balance_lik}
\textup{Pr}(\T \mid r)\propto \prod_{(j\to l) \in E_{\vec {\mathcal T}}} q_{j,l},
\]
corresponding to $\rho_r \propto 1$.
Therefore, we can simply draw the root from $\textup{Pr}(r)\propto g_r$, and use a cover algorithm with $W=Q$ to obtain samples distributed according to \eqref{eq:product_prob}. 

\noindent\textbf{Case 2, General form}: For a general and irreducible $Q\in [0,\infty)^{m\times m}$, the stationary distribution vector $\pi$ is a deterministic transform 
 of a left eigenvector of the transition probability matrix  $P$. Here, $({p_{j,l}\pi_j})\propto q_{j,l}$ may not hold for all $(j,l)$. To solve this problem, we introduce an auxiliary Markov chain as a mathematical tool for calculating $P$. Consider a chain $Y=(y_0,y_1,\ldots)$ over $V_G$ with transition kernel:
\(
\tilde p_{l,j}=\textup{Pr}(y_{t+1}=j \mid y_{t}=l)= \frac{q^*_{l,j}}{\sum_{k=1}^m q^*_{l,k}}, \qquad q^*_{l,j} = q_{j,l}.
\)
We denote the invariant distribution of $y_t:t\to \infty$ by $(\pi^*_1,\ldots, \pi^*_m)$, which can be numerically computed as a left eigenvector of the transition matrix formed by the $\tilde p_{l,j}$'s.

We now form the chain  $X=(x_0,x_1,\ldots)$  with the transition probability specified as:
\[\label{eq:forward}
p_{j,l}= \frac{ \tilde p_{l,j} \pi^*_l}{\pi^*_j}.
\]
Reversibility ($p_{j,l}=\tilde p_{j,l}$) is not needed; however, we can see that $\sum_j p_{j,l}\pi^*_j = \pi^*_l$ for all $l$. Therefore,  $X$ has the same invariant distribution $(\pi_1,\ldots, \pi_m)=(\pi^*_1,\ldots, \pi^*_m)$ as $Y$. As a result, we  can run a cover algorithm using $P$ from \eqref{eq:forward}. The probability \eqref{eq:ext_ab} in Theorem 1 becomes
\[
\textup{Pr}( \T \mid r) 
\propto \prod_{(j\to l) \in E_{\vec {\mathcal T}}}( \tilde p_{l,j} \pi^*_l)
=
\prod_{(j\to l) \in E_{\vec {\mathcal T}}} \frac{q_{j,l}}{\sum_{k=1}^m
q_{k,l}} \pi^*_l \propto  
\Big(
\prod_{(j\to l) \in E_{\vec {\mathcal T}}}q_{j,l}\Big)\Big( 
\frac{\sum_{k=1}^m
q_{k,r}}{\pi^*_r}
\Big),
\label{eq:r_cond}
\]
using the invariance to $\T$ implied by ${\pi^*_r}/({\sum_{k=1}^m
q_{k,r}})\prod_{(j\to l) \in E_{\vec {\mathcal T}}} {\pi^*_l}/({\sum_{k=1}^m
q_{k,l}})= \prod_{l=1}^m {\pi^*_l}/({\sum_{k=1}^m
q_{k,l}})$. 

The normalizing constant is now varying with $r$, $\rho_r \propto ( 
{\sum_{k=1}^m
q_{k,r}}/{\pi^*_r}
)$. Therefore, we draw the root from $\textup{Pr}(r)\propto g_r/\rho_r$, and then use a cover algorithm to draw from $\textup{Pr}(\T\mid r)$.

\section{Simulations}
To investigate the gain in computational efficiency of our fast-forwarded cover algorithm over competitors, we conduct several simulations in \texttt{R} and compare the run times of various spanning tree sampling algorithms on a machine equipped with an Apple M4 Max chip. 

\begin{figure}[H]
    \centering
    \begin{subfigure}[t]{0.32\textwidth}
        \centering
        \includegraphics[width=1\textwidth]{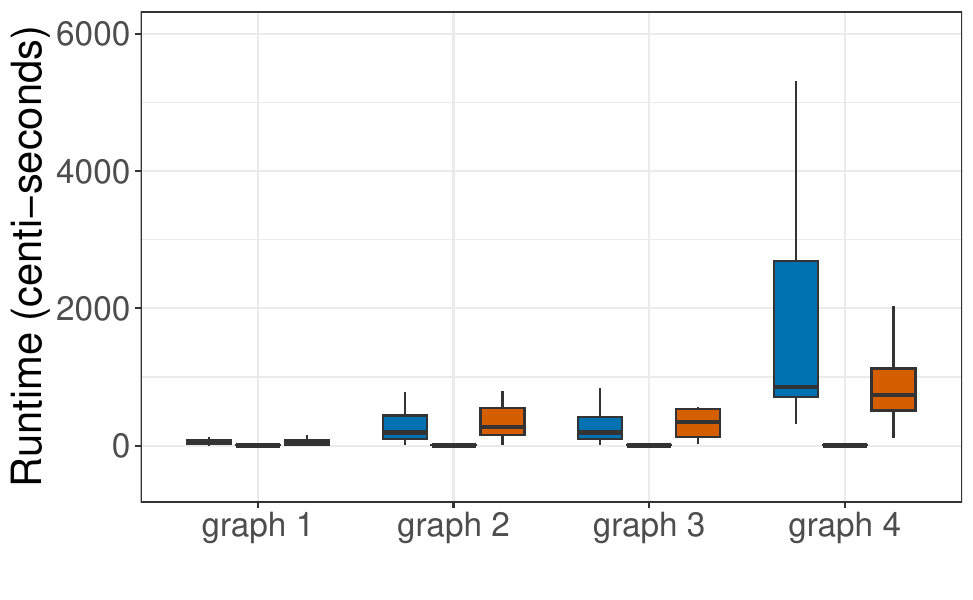}
        \caption{Runtime for different bottleneck size $1/\sqrt{\lambda_2(\mathcal L)}$.}
    \end{subfigure}\;
    \begin{subfigure}[t]{0.32\textwidth}
        \centering
        \includegraphics[width=1\textwidth]{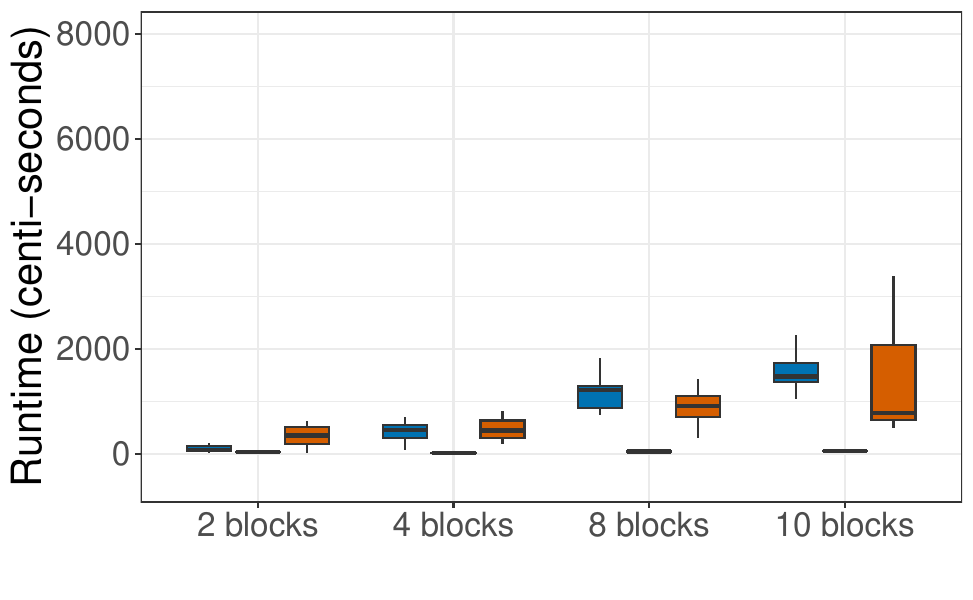}
        \caption{Runtime over different numbers of blocks and hence bottlenecks.}
    \end{subfigure}\;
    \begin{subfigure}[t]{0.32\textwidth}
        \centering
        \includegraphics[width=1\textwidth]{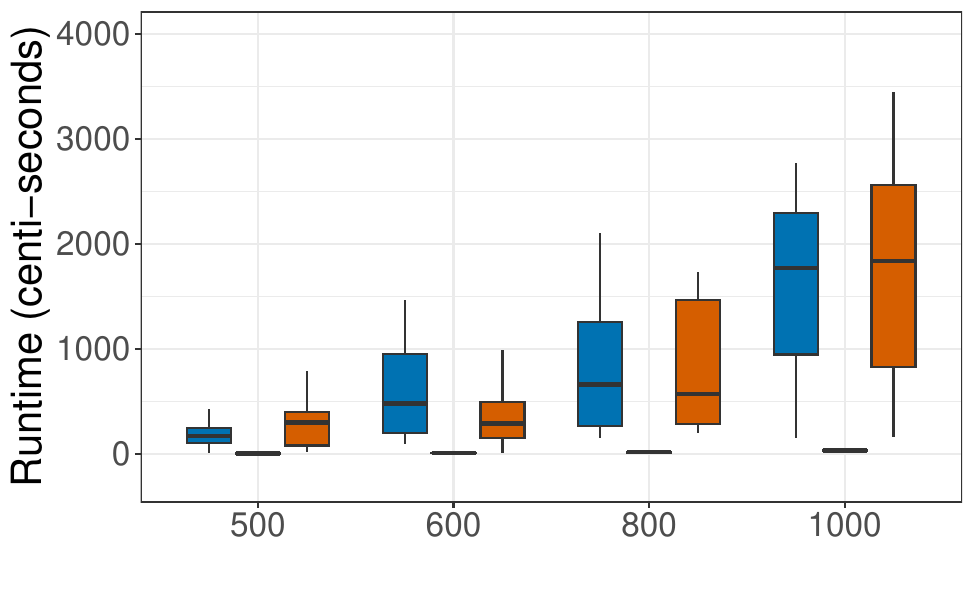}
        \caption{Runtime over different numbers of nodes.}
    \end{subfigure}
    \caption{Comparisons of runtime between Aldous--Broder algorithm (blue), fast-forwarded cover (grey) algorithm, and Wilson's algorithm (orange)}. In each setting, the fast-forwarded algorithm runtime has a small mean and variance, so that each grey box appears close to a thin line.    \label{fig:runtime}
\end{figure}

We compare the fast-forwarded cover algorithm against other random-walk-based competitors, namely the Aldous--Broder algorithm and Wilson's algorithm. First, we assess the impact of the bottleneck size, quantified via $1/\sqrt{\lambda_2(\mathcal L)}$, on wall-clock runtime. We simulate a graph with $m = 500$ nodes, where we partition the nodes $V$ into two blocks with $|V_1|=|V_2|=250$. The graph is represented by a symmetric weight matrix $W\in\mathbb{R}^{500\times 500}$ with each $w_{j,l}=u_{j,l}b_{j,l}$. The factor $u_{j,l}$ representing the size of the weight is simulated from $\text{Unif}(0,1)$. We set $b_{j,l}=|V_1|^2$ for those $j$ and $l$ that are in the same node partition, and $b_{j,l} \sim \text{Bern}(\zeta)$ for those $j$ and $l$ that are in different partitions. This creates two complete subgraphs connected by a small number of edges having small weight. We perform experiments at different edge densities with $\zeta$ from $(0.5,0.1,0.05,0.01)$, corresponding to a range of bottleneck values near $1/\sqrt{\lambda_2(\mathcal L)}=(249, 560, 786, 1738)$, corresponding to graphs 1 to 4 in Figure \ref{fig:runtime}(a). Under each value of $\zeta$, we draw 10 spanning trees from each of the three algorithms under comparison. Figure \ref{fig:runtime}(a) shows that while the runtimes for Aldous--Broder and Wilson's algorithm increase rapidly as the bottleneck size increases, the fast-forwarded algorithm runtime is almost unaffected by bottleneck size. Similar plots are shown in the supplementary materials, where we compare the Aldous--Broder algorithm, Wilson's algorithm and the fast-forwarded algorithm under the same setting but measure time by number of random walk steps taken. 

Second, we assess the effects of the number of blocks on runtime; more blocks implies more bottlenecks. We simulate $W\in\mathbb{R}^{600\times 600}$ according to $w_{j,l}=u_{j,l}b_{j,l}c_{j,l}$. The factors $u_{j,l}$ and $b_{j,l}$ are simulated as above, except that we partition the $600$ nodes into $K$ blocks $(V_1,\ldots, V_K)$ each of node size $(600/K)$, and use $b_{j,l}=b_{l,j}\sim \text{Bern}(0.001)$ and $c_{j,l}=0.005/K$ for those $(j,l):j\in V_k, l \in V \setminus V_{k}$. The factors $c_{j,l}$ are chosen so that empirically the size of $1/\sqrt{\lambda_2(\mathcal L)}$ remains roughly the same.  We vary $K$ in $(2,4,8,10)$, and draw 10 spanning trees using each of the algorithms under comparison. Figure \ref{fig:runtime}(b) shows that runtimes of the Aldous--Broder and Wilson's algorithms increase quickly as the number of blocks increases, while the fast-forwarded algorithm is again almost unaffected.

Third, we assess scalability by running the samplers on graphs with numbers of nodes ranging within $(500,600,800,1000)$ in the two-block case using $b_{j,l}=b_{l,j}\sim \text{Bern}(0.1)$. Figure \ref{fig:runtime}(c) shows that the runtimes for the Aldous--Broder and Wilson's  algorithms increase much faster than the runtime of the fast-forwarded cover algorithm.

In the Supplementary Materials, we present a comparison with a Laplacian-based sampler outlined in Algorithm 1 of \cite{harvey2016Generating}.

\section{Bayesian Dendrogram Inference for Crime and Communities Data}\label{sec:dendrogram}

We demonstrate the use of our algorithm in inferring a Bayesian dendrogram. We consider a hierarchical model in the form of \eqref{eq:BayesDendro} for continuous $y_i\in \mathbb{R}^{\tilde d}$ with $\T$ rooted at node $1$. We choose 
\(
 \mathcal F (y_i \mid \mu_{z_i}) = \phi(y_i;\mu_{z_i}, \Sigma),  \qquad \mathcal H(\mu_l\mid \mu_j, \Sigma) = \phi(\mu_l; \mu_j,\lambda^{-1} \Sigma).
\)
By symmetry of the kernel $\mathcal{H}$, the underlying graph $G$ is undirected. This leads to the following hierarchical model,
\(L(y \mid \mu, z)  &= \prod_{i=1}^n\phi(y_i;\mu_{z_i}, \Sigma),\qquad
 \Pi(\mu \mid E_{\vec{\mathcal{T}}}, r)  &= R(\mu_r) \prod_{(j\to l)\in E_{\vec{\mathcal{T}}}} \phi(\mu_l; \mu_j,\lambda^{-1} \Sigma)
 \) with priors 
$\Pi_0(\vec{\mathcal{T}}\mid r=1) \propto 1$, $z_i \stackrel{iid} \sim \text{Categorical}({\tilde \pi}_1,\ldots, {\tilde \pi}_{\tilde K} )$, $({\tilde \pi}_1,\ldots, {\tilde \pi}_{\tilde K} )\sim \text{Dir}(\alpha,\ldots, \alpha)$, $\Sigma \sim \text{W}^{-1}(\nu, \Sigma_0)$, $R(\mu_1 = 0) = 1$. Since
$\mathcal{F}$ and $\mathcal{H}$ have a shared covariance $\Sigma$ up to a scale change $\lambda>0$, a conjugate normal-inverse Wishart prior can be chosen for
$\mu$ and $\Sigma$. 

Our interest is inferring the posterior on $\vec T$, but we also have uncertainty in allocations $z_i$ and component-specific parameters $(\mu_k)$. The resulting joint posterior distribution is highly complex. Due to computational challenges, the literature tends to avoid characterizing the uncertainty in
$\vec T$. For example, for Bayesian hierarchical clustering, \cite{heller2005bayesian} uses a hypothesis testing criterion to iteratively merge clusters, while \cite{heard2006quantitative} combines clusters based on a metric that captures the closeness between clusters. Alternatively, one can use a Markov chain Monte Carlo sampling algorithm. Classical algorithms rely on making local changes in $\vec T$ using reversible jump Metropolis-Hastings \citep{chipman1998bayesian,denison1998bayesian}, which tends to be very inefficient. \cite{wu2007bayesian} considers adding a move that allows a larger tree to be restructured to improve mixing, but with high per iteration expense for large graphs.

Our new spanning tree sampler can bypass these computational challenges. We take an overfitted modeling approach, considering an encompassing tree $\T$ with $\tilde m$ nodes rooted at 1, with $\tilde m$ sufficiently large to provide an upper bound on the true value $\tilde m\ge m$. This allows us to build a blocked Gibbs sampler based on drawing from $\Pi(z \mid y, \mu, \T)$, $\Pi(\mu \mid y, z, \T)$ and $\Pi(\T \mid \mu )$, with $\T$ a spanning tree for $\tilde m$ nodes, in addition to the steps of updating other parameters.

After obtaining a posterior sample, we can marginalize redundant nodes and change each sampled spanning tree $\T$ to a reduced dendrogram $\vec T$, while maintaining an equivalent generative model for the data. Given a sample of $(z_1,\ldots, z_n)$, and $\vec T$ initialized at $\T$, we use the following pruning procedures corresponding to integrating out the densities related to $j$:
\begin{itemize}
	\item If $j$ is an empty leaf node, without any downstream edge $(j\to l)$ and with $\sum_{i=1}^n 1(z_i=j)=n_j=0$, we remove $j$ and $(k\to j)$ from $\vec T$.
	\item If $j$ only has two edges $(k\to j)$ and $(j\to l)$, and $n_j=0$, we remove $j$ and replace the two edges by $(k\to l)$ in $\vec T$.
\end{itemize}
We iterate the above pruning steps on the tree's nodes until we cannot reduce the size of $\vec T$ any further. The spanning tree $\T$ can be viewed as a latent variable that facilitates the specification of the model and posterior computation for the dendrogram $\vec T$. We refer to this approach as a {\em spanning tree-augmented dendrogram.} %(STAD)

To illustrate this model, we consider an application to the Community and Crimes dataset from the University of California at Irvine Machine Learning Repository. The dataset consists of socioeconomic attributes from the 1990 United States Census, crime statistics from the 1995 Uniform Crime Report, and law enforcement attributes from the 1990 Law Enforcement Management and Administrative Statistics Survey.  We focus on economic data from the state of Massachusetts, where there are $n=123$ relevant entries, each corresponding to a community in the state, with ${\tilde d}=2$ continuous attributes: the community's median income and median rent. These attributes are log-transformed and standardized to have sample mean $0$ and marginal sample variance $1$. Our focus is on characterizing variability in these economic attributes by grouping the communities hierarchically. A dendrogram is natural for this purpose. 

We choose priors to favor node parameters $\mu_j$ that are broadly spread across the support of the data, with relatively few data points associated with each $\mu_j$. This is achieved by choosing $\nu = n$,  $\Sigma_0 = 0.2^2 I_{\tilde d}$, $\lambda = 0.25$ and $\tilde m = \left \lfloor{n/4}\right \rfloor $. To favor effective elimination of unnecessary clusters, we follow common practice in the literature on overfitted mixtures \citep{van2015overfitting} and choose a symmetric Dirichlet with a small concentration parameter ($\alpha=0.1$) for weights ${\tilde \pi}_1,\ldots,{\tilde \pi}_{\tilde{m}}$. 
Since the data have been centered, we fix root choice $r=1$ and $\mu_1=0$. There are implicit Bayesian Occam's razor effects that favor the induced dendrogram $\vec T$ to be small. As in other Bayesian mixture models, the marginal likelihood will tend to decrease if data are over clustered, favoring setting $n_k=0$ to automatically remove some of the clusters. This tendency is furthered by our symmetric Dirichlet prior with precision close to zero for the cluster weights. In addition, the uniform prior on $\T$ leads to higher weight on dendrograms with few nodes as such dendrograms have more ways to be marginalized (pruned) from a $\tilde m$-node $\vec T$.  

Using a Gibbs sampler, we can update each term above using closed-form full conditional distributions. We provide the details in the supplementary material. To show computational advantages of this Gibbs sampler, we compare with sampling the posterior for a directly specified dendrogram using a reversible jump Markov chain Monte Carlo sampler. The model is almost the same as our spanning tree-augmented version, with the same choice of $\mathcal F$ and $\mathcal G$, priors for the parameters, and upper bound $\tilde m$ on the number of nodes in $\vec T$. However, we allow the number of nodes in $\vec T$ to vary; therefore, some nodes amongst $(1,\ldots, \tilde m)$ might not be in $\vec T$. In the likelihood, we replace $ \mathcal F(y_i \mid \mu_{k}) $ by zero if $k$ is not a node of $\vec T$, and use the prior $\Pi_0(\vec T) \propto 0.01^{|V_{\vec T}|}$ to favor small dendrograms. Accordingly, we use birth/death proposals to add/remove nodes from $\vec T$, and a Metropolis-Hastings criterion to accept or reject each proposal. We provide details in the supplementary material.

\begin{figure}[H]
\begin{subfigure}[t]{.24\textwidth}
    \begin{overpic}[width=\textwidth]{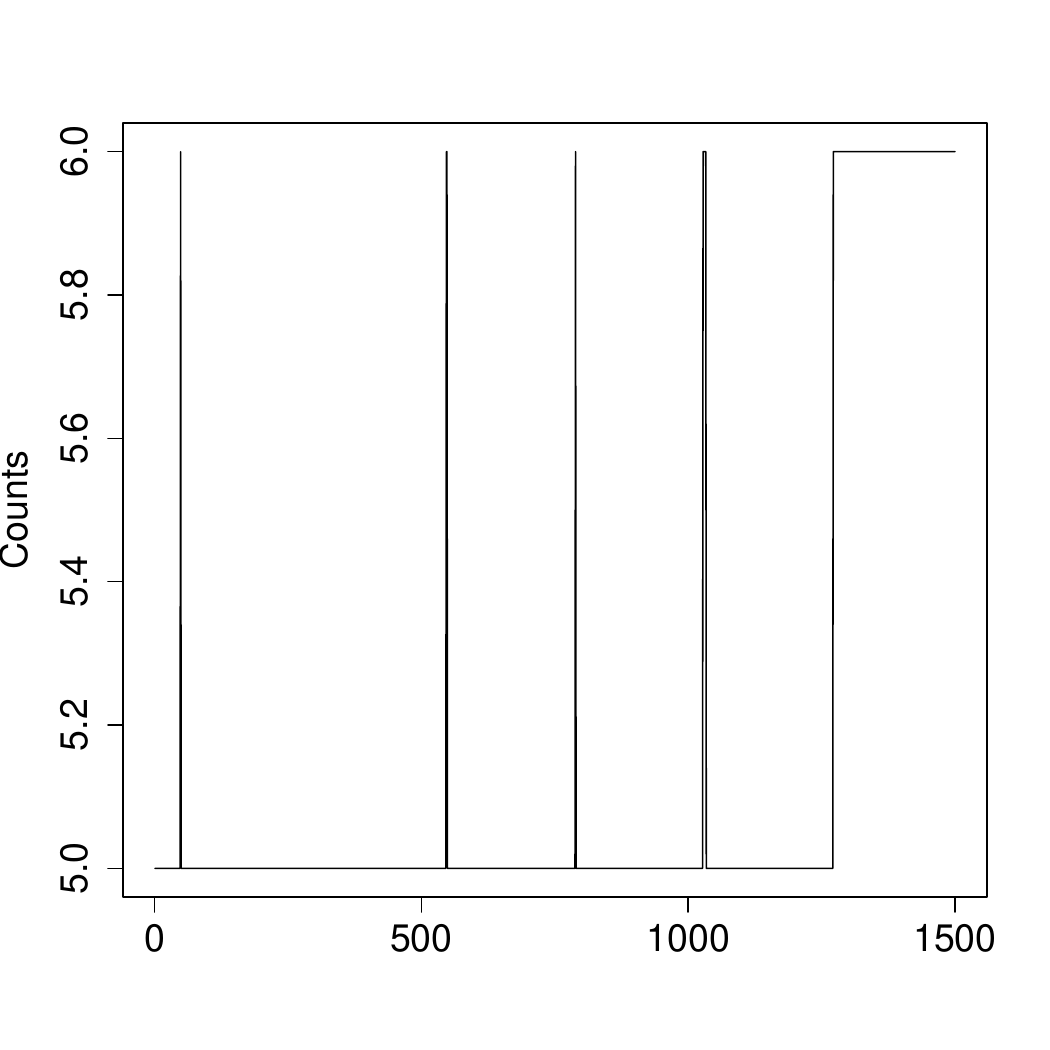}
    \put(45, -1){\scriptsize Iteration}
    \end{overpic}
    \caption{\scriptsize Trace of maximum degree  from reversible jump sampler.}
\end{subfigure}
\begin{subfigure}[t]{.24\textwidth}
    \begin{overpic}[width=\textwidth]{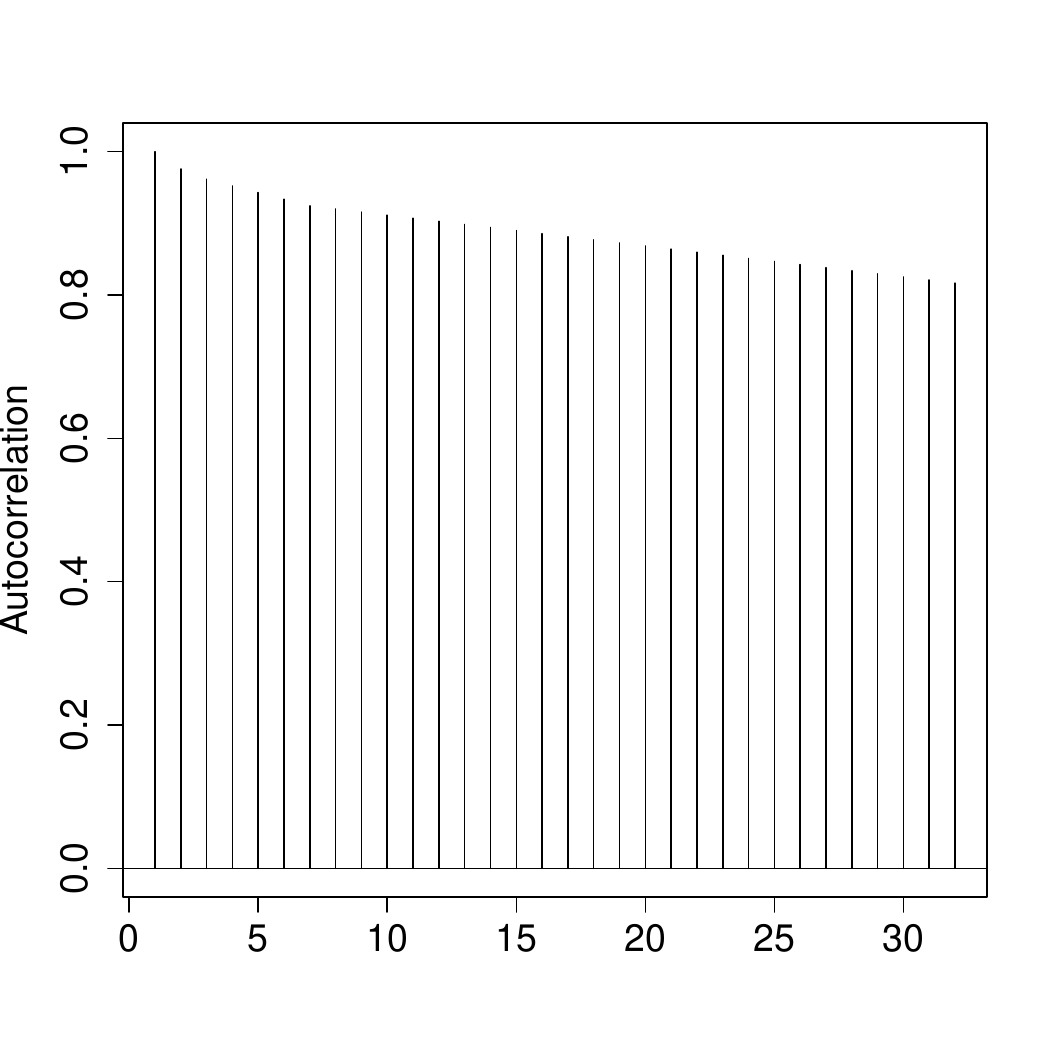}
    \put(50, -1){\scriptsize Lag}
    \end{overpic}
    \caption{\scriptsize Autocorrelation of maximum degree for reversible jump sampler.}
\end{subfigure}
\begin{subfigure}[t]{.24\textwidth}
    \begin{overpic}[width=\textwidth]{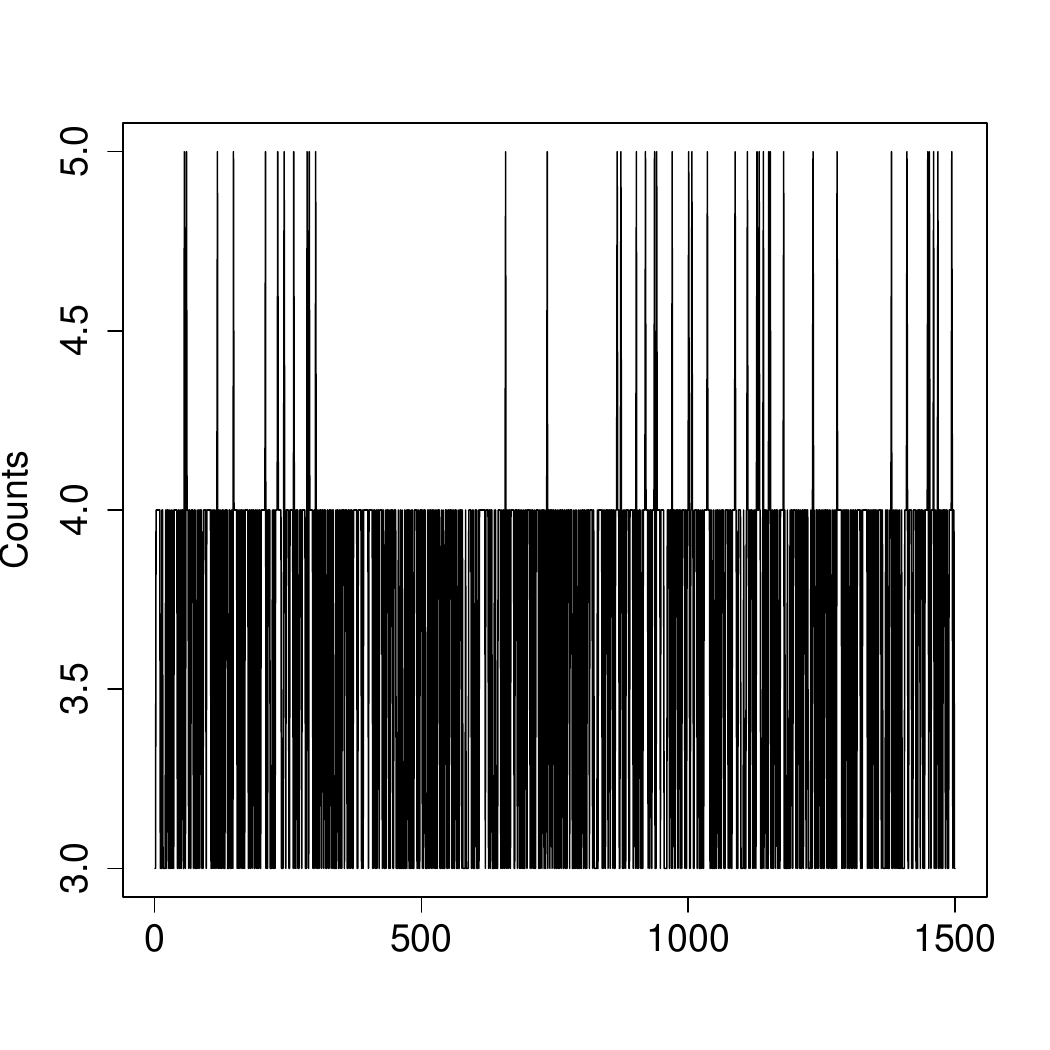}
    \put(45, -1){\scriptsize Iteration}
    \end{overpic}
    \caption{\scriptsize Trace of maximum degree from Gibbs sampler.}
\end{subfigure}
\begin{subfigure}[t]{.24\textwidth}
    \begin{overpic}[width=\textwidth]{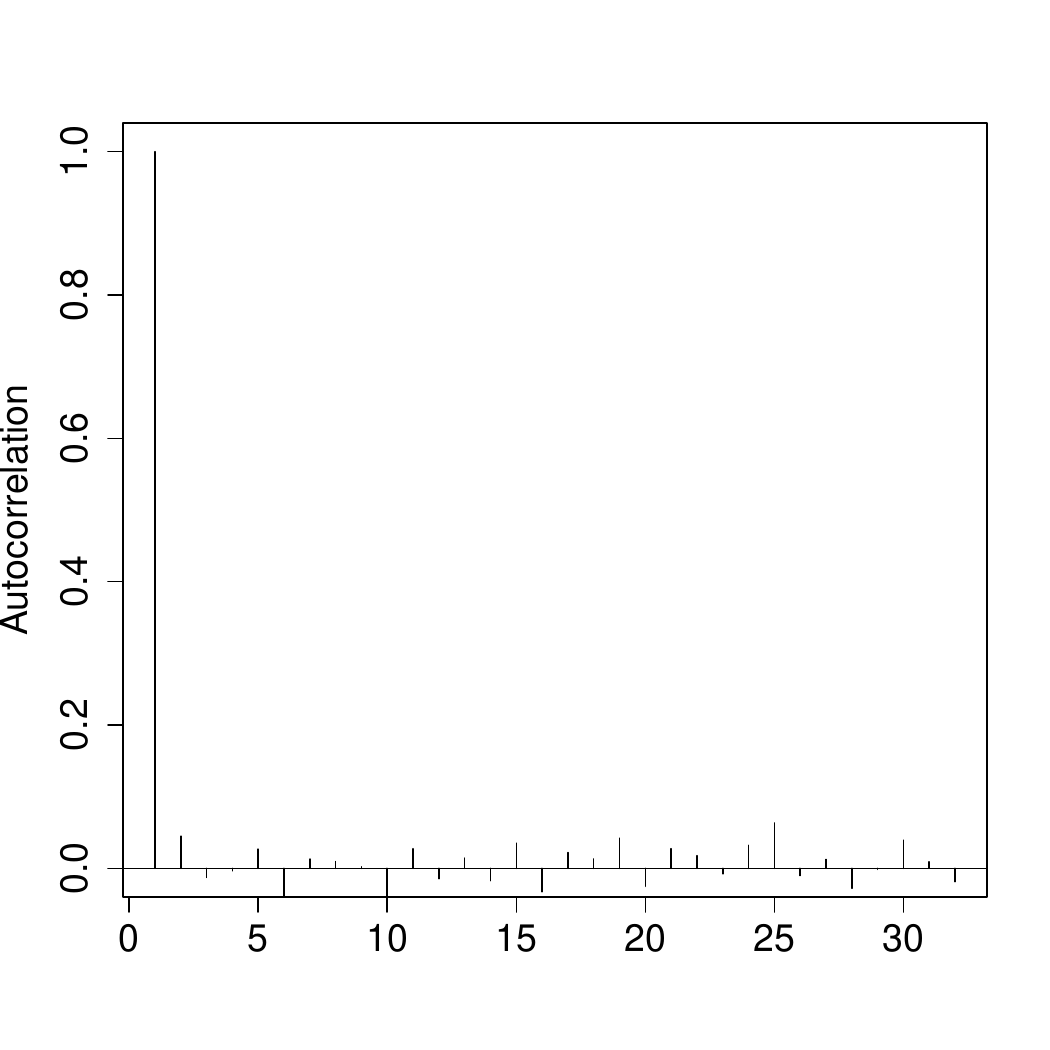}
    \put(50, -1){\scriptsize Lag}
    \end{overpic}
    \caption{\scriptsize Autocorrelation of maximum degree  from Gibbs sampler.}
\end{subfigure}
\begin{subfigure}[t]{.24\textwidth}
    \begin{overpic}[width=\textwidth]{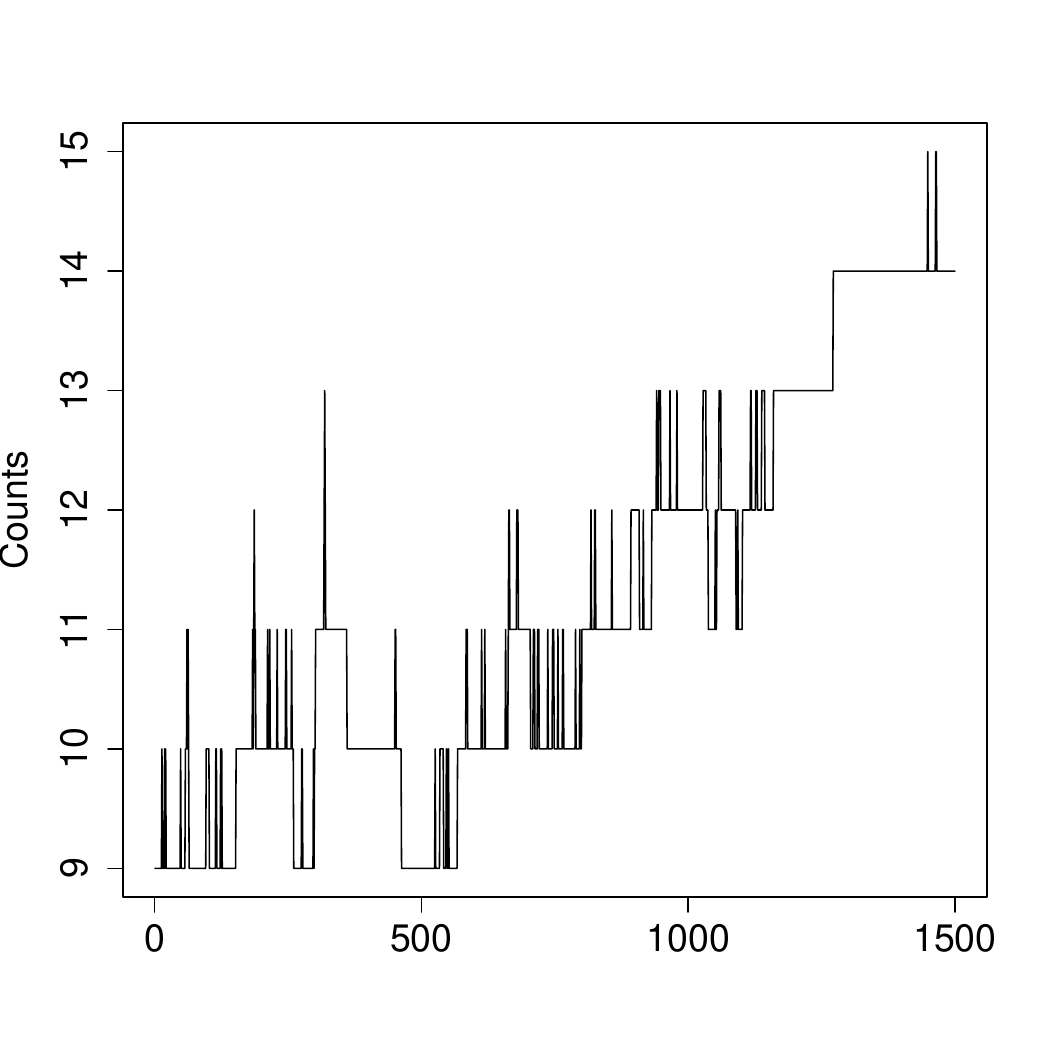}
    \put(45, -1){\scriptsize Iteration}
    \end{overpic}
    \caption{\scriptsize Trace of number of non-empty  leaves for reversible jump sampler.}
\end{subfigure}
\;
\begin{subfigure}[t]{.24\textwidth}
    \begin{overpic}[width=\textwidth]{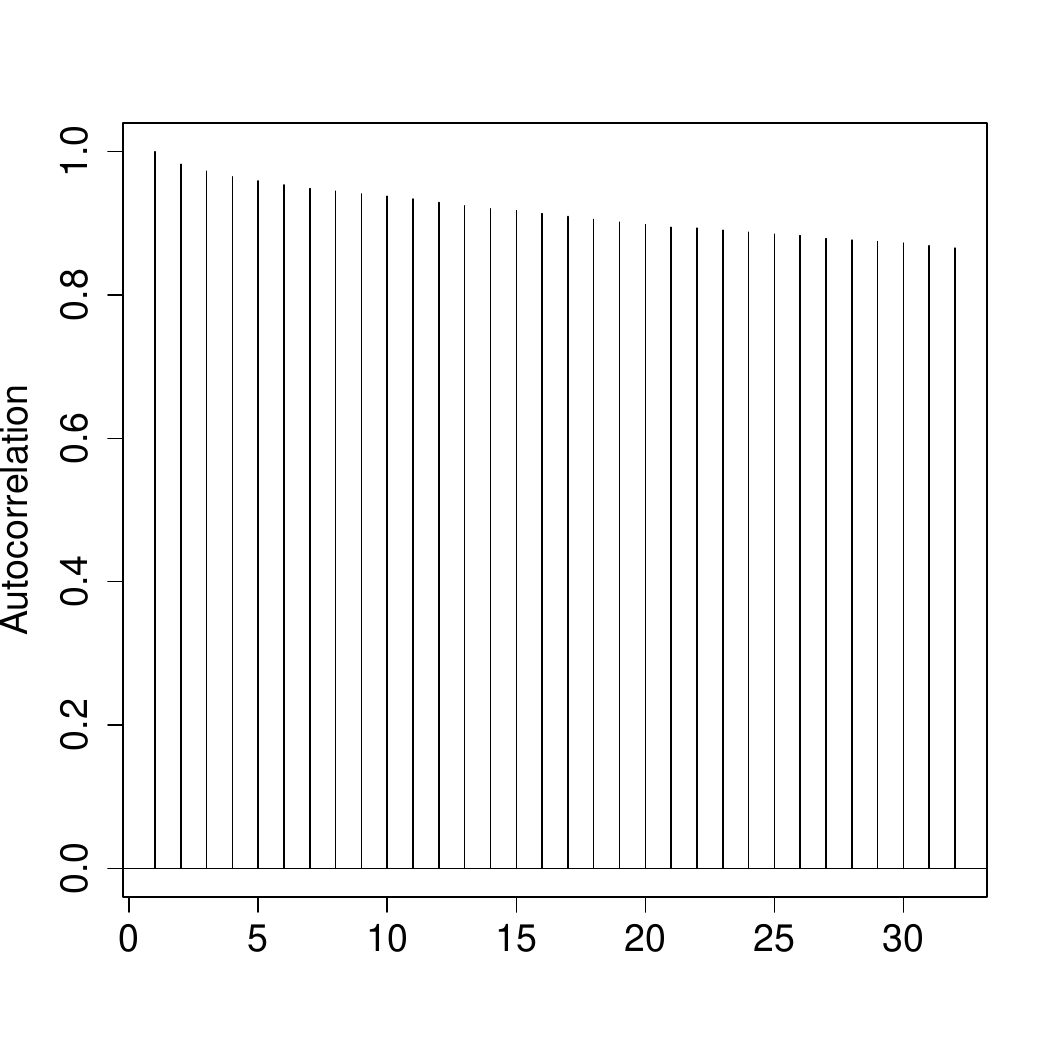}
    \put(50, -1){\scriptsize Lag}
    \end{overpic}
    \caption{\scriptsize Autocorrelation of number of non-empty  leaves for reversible jump sampler. }
\end{subfigure}
\;
\begin{subfigure}[t]{.24\textwidth}
    \begin{overpic}[width=\textwidth]{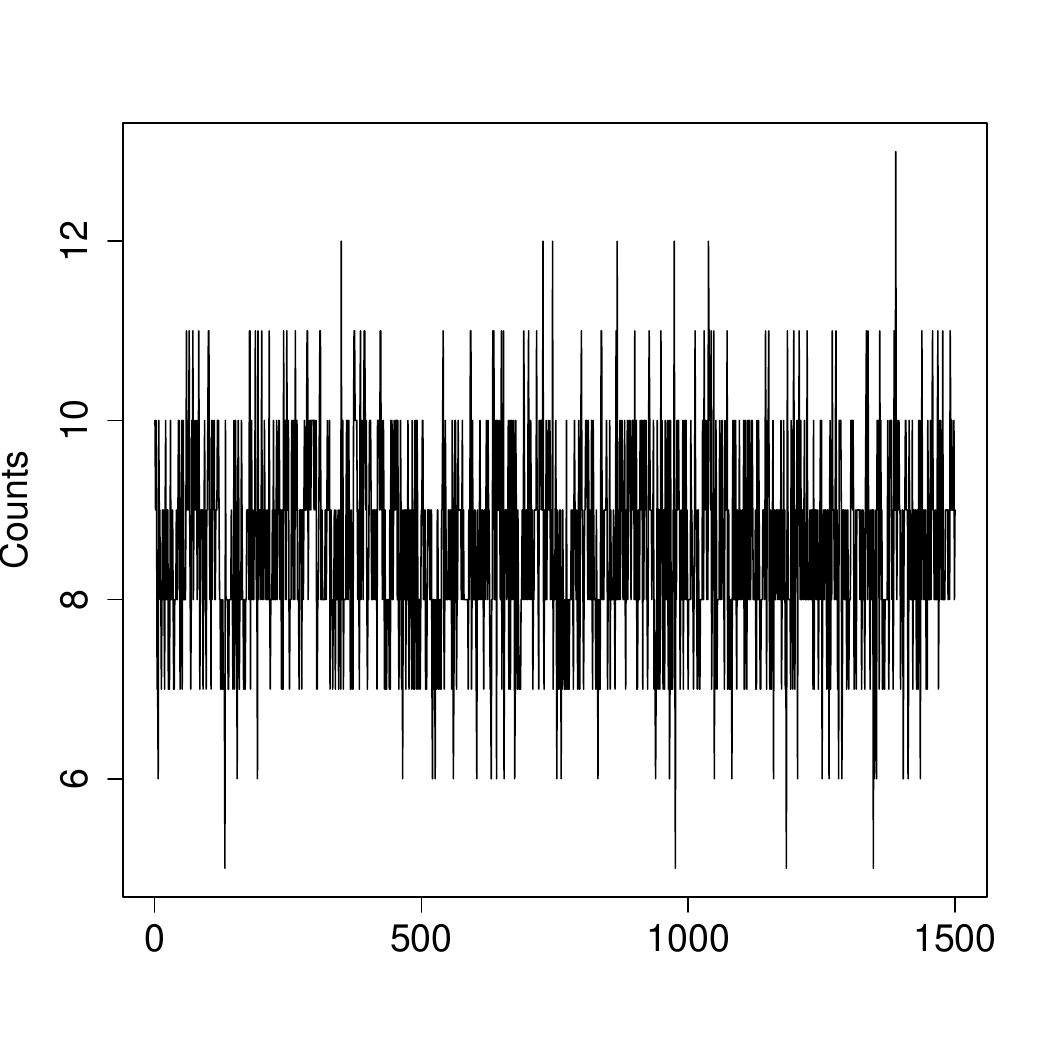}
    \put(45, -1){\scriptsize Iteration}
    \end{overpic}
    \caption{\scriptsize Trace of number of non-empty  leaves from Gibbs sampler.}
\end{subfigure}
\;
\begin{subfigure}[t]{.22\textwidth}
    \begin{overpic}[width=\textwidth]{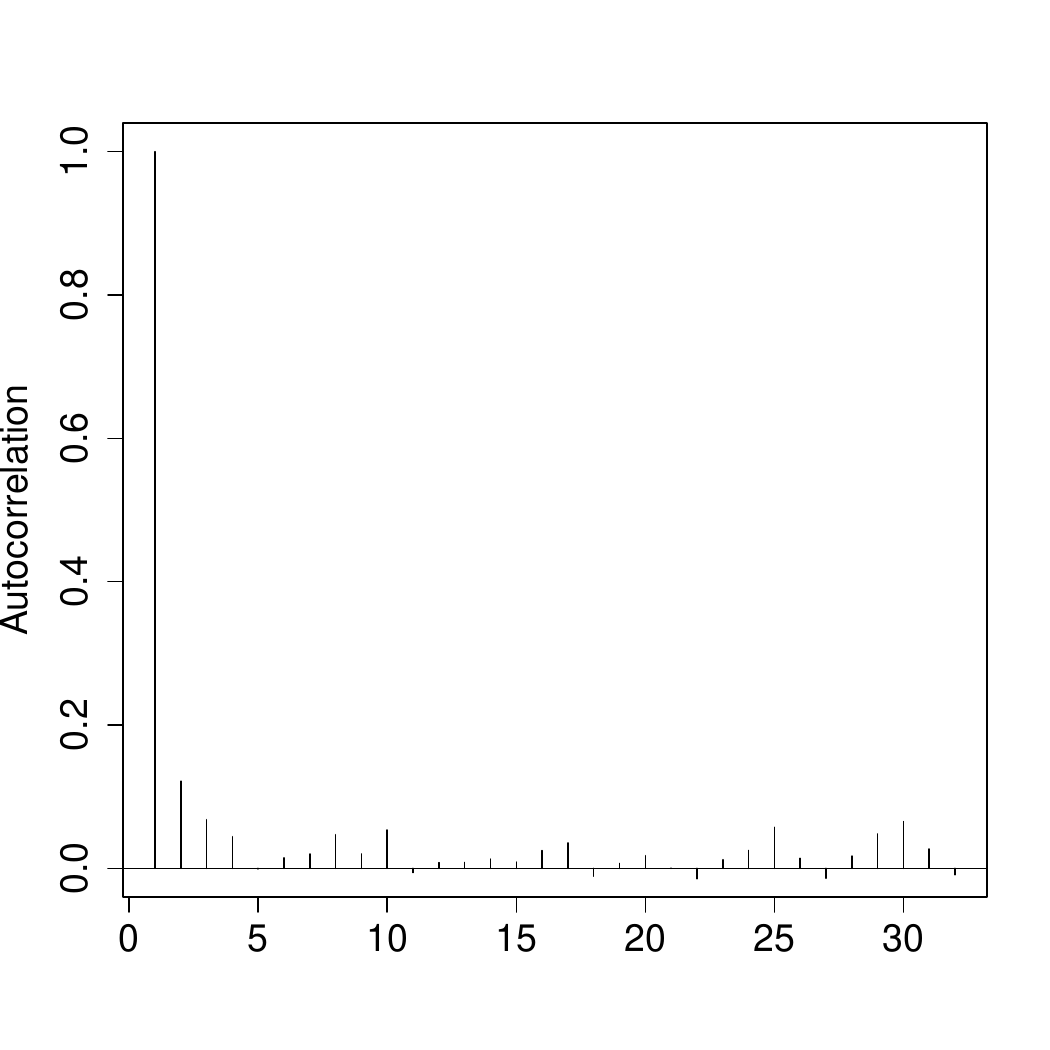}
    \put(50, -1){\scriptsize Lag}
    \end{overpic}
    \caption{\scriptsize Autocorrelation of number of non-empty  leaves from Gibbs sampler.}
\end{subfigure}
\caption{The Gibbs sampler for a spanning tree-augmented dendrogram model shows much faster mixing, compared to a reversible jump sampler for a directly specified dendrogram model. The posterior distributions targeted by the Gibbs sampler and reversible jump sampler are slightly different due to how the complexity of tree is regulated.}\label{fig:gibbs}
\end{figure}

We run each sampler for $5000$ iterations, with a burn-in of $3500$ iterations. In wall clock time, both the Gibbs sampler using our fast-forward cover algorithm and the reversible jump Markov chain Monte Carlo sampler run for approximately $0.5$ to $1$ hours. However, there are dramatic differences in mixing performance. The reversible jump Markov chain Monte Carlo sampler tends to be stuck in certain states for a long time, while the spanning tree-based Gibbs sampler shows excellent mixing. Figure \ref{fig:gibbs} compares the mixing for the number of nonempty leaves of the dendrogram, with the results for other summaries shown in the Supplementary Materials. We compare effective sample sizes per iteration in Table \ref{ess}.
We further conducted posterior sampling using the subtree prune and regraft move \citep{evans2005subtree, song2006properties}. The results are provided in the Supplementary Materials.

\begin{figure}[H]

\begin{subfigure}[t]{.3\textwidth}
  \includegraphics[width=\textwidth]{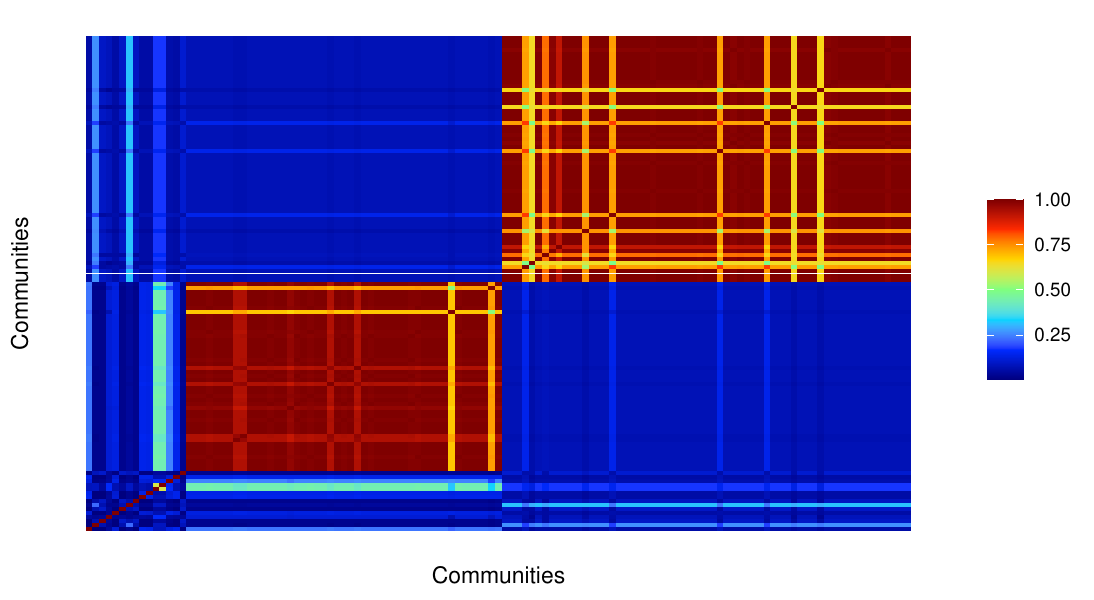}
\caption{Similarity matrix at depth 1.}
\end{subfigure}
\;
\begin{subfigure}[t]{.3\textwidth}
  \includegraphics[width=\textwidth]{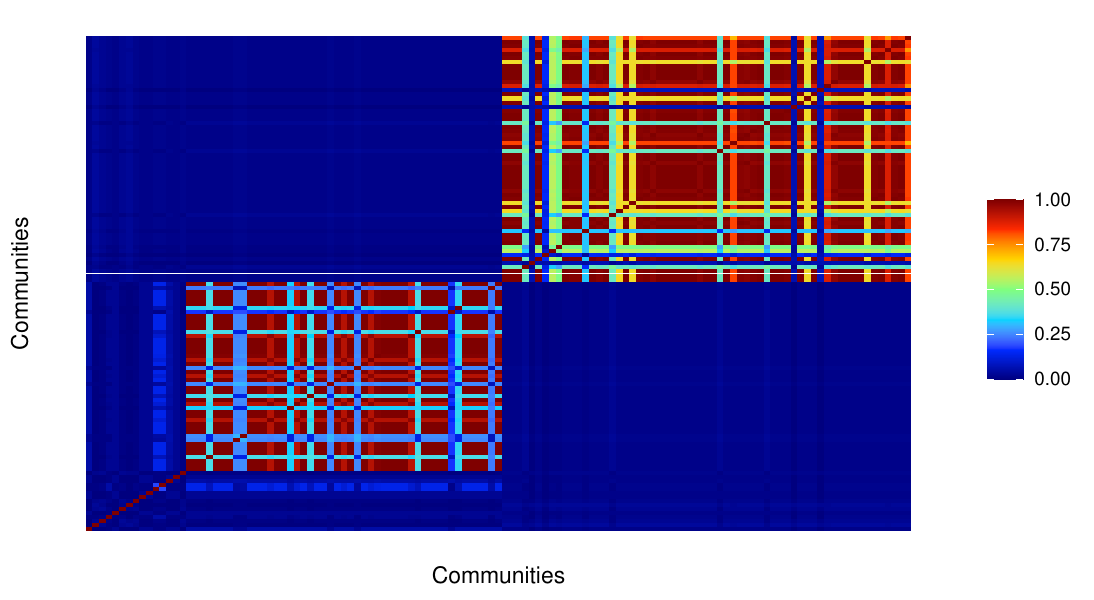}
\caption{Similarity matrix at depth 2.}
\end{subfigure}
\;
\begin{subfigure}[t]{.3\textwidth}
    \includegraphics[width=\textwidth]{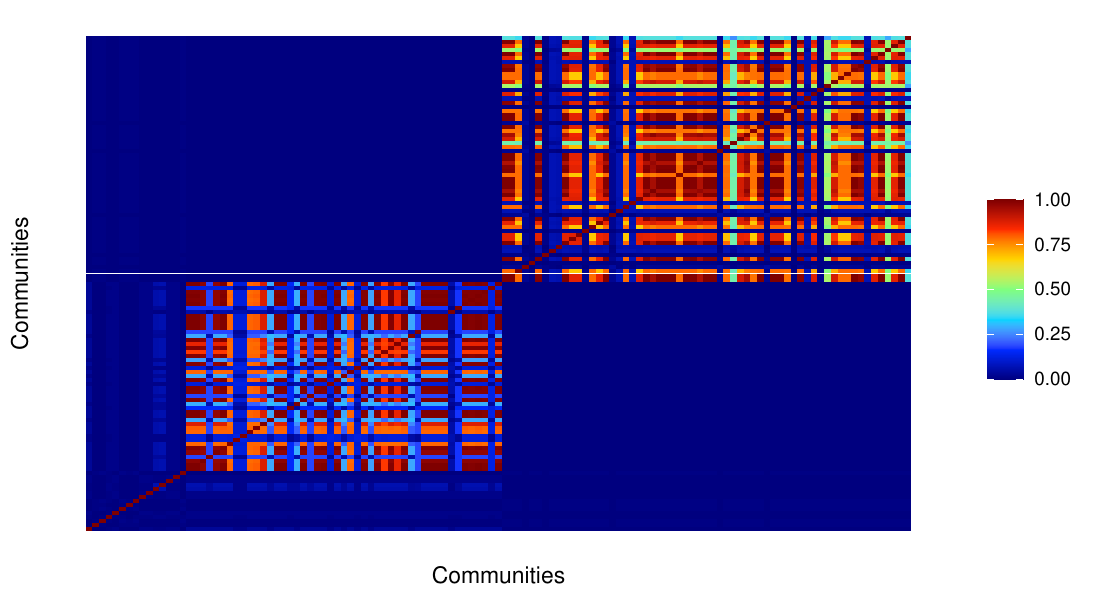}
    \caption{ Similarity matrix at depth 3.}
\end{subfigure}
\;
\caption{Posterior similarity matrices at different depths for crime and communities data\label{fig:similarity}.}

\end{figure}

To quantify the uncertainty around the obtained clusters and visualize the inferred hierarchical structures, we use the posterior similarity matrix \citep{fritsch2009improved}.
We record whether communities share an ancestor node at depth $1$, $2$ and $3$ of the sampled dendrogram, and average over the posterior samples to compute a probability for each such pairing. The results are shown in heat maps in Figure \ref{fig:similarity}. One can observe clear block-diagonal structures at all depths, which become increasingly noisy as the depth increases.

\begin{figure}[H]
\begin{subfigure}[t]{.5\textwidth}
\includegraphics[width=\textwidth]{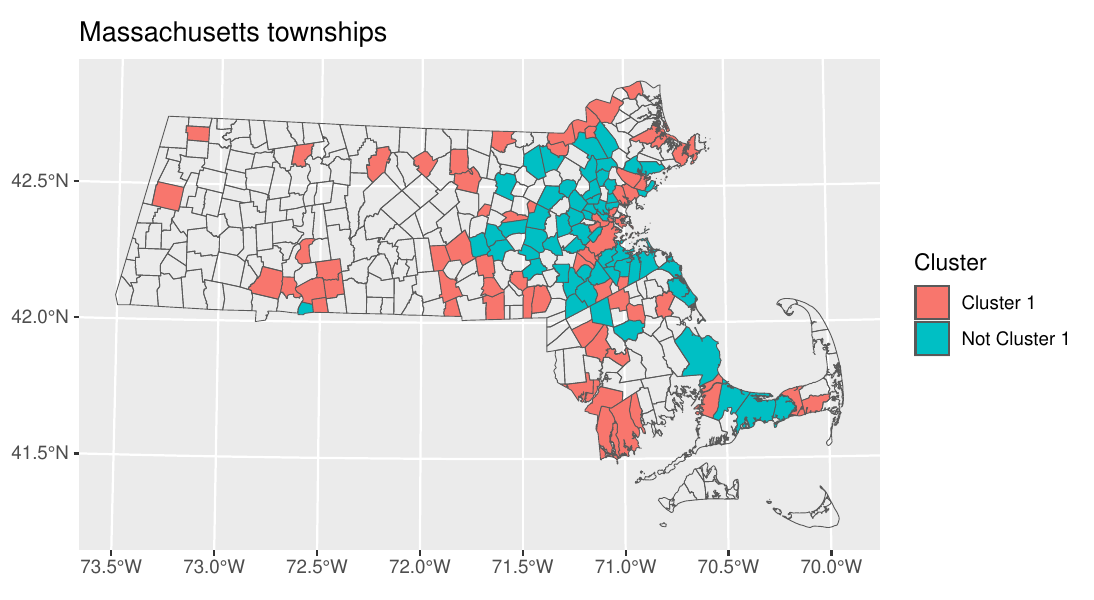}
\caption{Map plot of Massachusetts communities, colored membership of a depth 1 cluster.}
\end{subfigure}
\;
\begin{subfigure}[t]{.5\textwidth}
    \includegraphics[width=\textwidth]{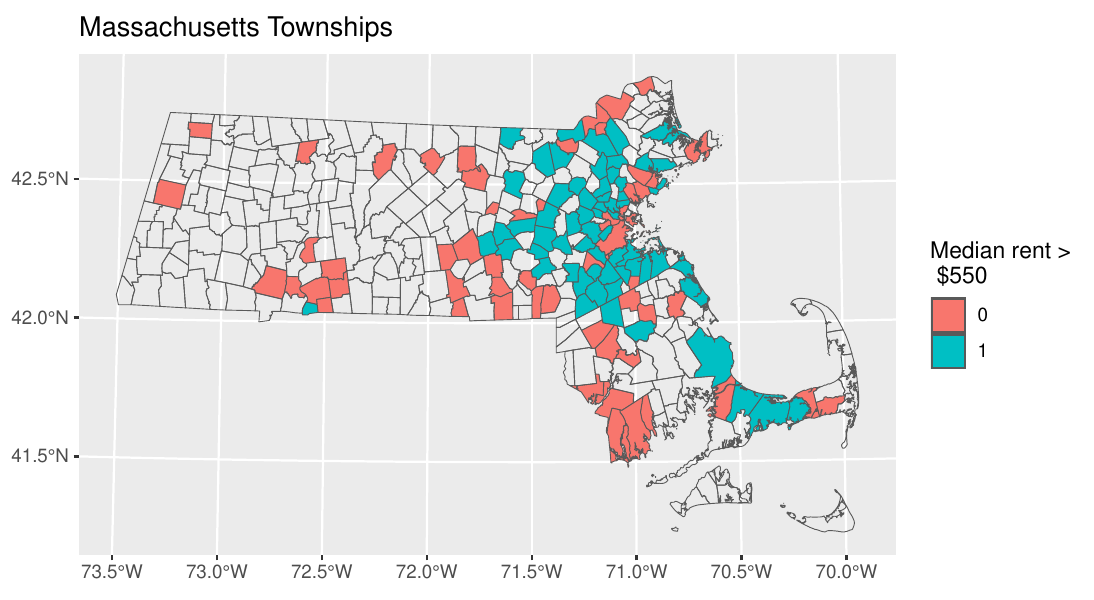}
    \caption{ Map plot of Massachusetts communities, colored by whether median rent is above threshold.}
\end{subfigure}
\caption{The communities in the largest cluster, red in panel (a), at depth $1$ of the estimated dendrogram roughly match with those with median rent above $\$550$ in panel (b)\label{fig:map}.}

\end{figure}

To interpret the inferred clusters, we visualize the communities in Massachusetts on a map. We identified the largest diagonal block obtained from the posterior similarity matrix at depth $1$ and colored the map by membership. We juxtapose the cluster membership map with another map colored by whether the median rent in the community is above the threshold of $\$550$ in Figure \ref{fig:map}. There is a nearly identical correspondence between the two maps, suggesting that the clustering faithfully captures the variation in the data. We also observed a visible concentration of such higher income and rent communities in eastern Massachusetts near the coast.  Details for computing posterior similarity matrices and maps are provided in the Supplementary Materials. 

\begin{table}
\def~{\hphantom{0}}
\tbl{Effective sample size per  iteration for the inferred tree 
from the reversible jump sampler and our proposed spanning tree-augmented dendrogram Gibbs sampler.}{%
\begin{tabular}{lcc}
\hline
Parameters & Gibbs sampler for spanning tree-augmented dendrogram  & Reversible-jump sampler  \\[5pt]
\hline
Maximum degree      & 0.913        &    0.003\\
Maximum depth       & 0.29    &    0.007\\
Number of leaves    & 0.702    &    0.002\\
\hline
\end{tabular}}
\label{ess}
\end{table}

\section{Discussion}

This article is motivated by improving the computational efficiency of posterior inference for tree parameters. Expanding beyond the application to dendrograms, it is of interest to develop spanning tree-augmented models for more complex models, 
such as more elaborate discrete latent structure models \citep{zeng2023tensor} or treed Gaussian process \citep{gramacy2008bayesian,payne2024bayesian}. For broader classes of tree models that may not admit a product-over-edges probability distribution, one may modify our random walk cover algorithm to form a computationally efficient proposal-generating distribution. Related Metropolis-Hastings algorithms for sampling graph partitions have recently been studied by \cite{autry2023metropolized}. For a tree with unknown number of nodes or structural dependence on a latent arrival process, such as diffusion trees \citep{neal2003density} or Bayesian phylogenetic trees \citep{huelsenbeck2001mrbayes}, it is interesting to explore extending the fast-forwarding algorithm  to bypass wasteful random walks on an infinite graph or under time-varying transition probabilities.

\section*{Acknowledgement}
This work was partially supported by Merck, the European Research Council, the Office of Naval Research, the National Institutes of Health and the National Science Foundation of the United States.
\appendix

\appendixone
\section*{Appendix}

\subsection*{Code}
All code for this paper can be found at \url{www.github.com/edrictam/fast_spanning_tree_sampler}
\subsection*{Proofs}

\subsection*{Proof for Theorem 2}
\begin{proof}
Considering $(\hat t_{k+1} - \hat t_k)$ as the inter-arrival time, we want to show that
$$ \mathbb{E}(\hat t_{k+1} - \hat t_k) \ge 1/\tilde q, \qquad \tilde  q= \{1 + \min_{j\in U_k}\frac {\sum_{l\in  U_k} w_{j,l} } {\sum_{l'\in \bar U_k} w_{j,l'}}\}^{-1}.$$ 

Consider a sequence of independent Bernoulli events with success probabilities $\tilde{p}_1,\tilde{p}_2,\ldots$, with all $\tilde{p}_{\tilde{i}}\le \tilde{q}$, and denote the index on the first success in this sequence by $\tilde{T}$. If $\mathbb E \tilde{T}<\infty$ then $\mathbb E \tilde{T} \ge 1/\tilde{q}$. Since $\tilde{T}\ge 0$, we know
$\mathbb{E} \tilde{T}=\sum_{t=0}^{\infty} \text{Pr}(\tilde{T}> t) = \sum_{t=1}^{\infty} \prod_{\tilde{i}=1}^t (1-\tilde{p}_{\tilde{i}})
\ge  \sum_{t=1}^{\infty} \prod_{\tilde{i}=1}^t (1-\tilde q)=1/\tilde q$. 
Adding over $\hat t=\sum_{k=1}^{m} (\hat t_{k+1} - \hat t_k)$ with $\hat t_1=0$ yields the result.
\end{proof}

\subsection*{Proof for Theorem 3}

\begin{proof}
We first state Cheeger's inequality for circulation graphs [\cite{chung2005laplacians}, Theorem 5.1].
For a directed and weighted graph with non-negative weight matrix $W$, having $\sum_{l=1}^m w_{j,l}=\sum_{l=1}^m w_{l,j}$ for all $j$, the second smallest eigenvalue $\lambda_2(\mathcal L)$ satisfies
\(
\sqrt{\lambda_2(\mathcal L)} \ge \min_{U: 1\le |U|\le m-1  } \frac{\sum_{j\in U, l\in \bar U} w_{j,l}}{\min (\sum_{j\in U} d_j, \sum_{j\in \bar U} d_j)}.
\)
Letting $U= \cup_{\tilde t\le t} X_{\tilde t}$, $|U|\le m-1$, the probability of exiting $U$ at time $t+1$ is
\(
\text{Pr}(x_{t+1}\in \bar U \mid x_{\tilde t}\in U \;\forall \tilde{t} \le t) & = 
\sum_{j\in U} \bigg \{ \frac{\sum_{l\in \bar U} w_{j,l}}{d_j} S_t(j) \bigg\}.
\)
where $S_t(j)=\text{Pr}(x_t=j)$.
Using $\tilde e_j = \sum_{l\in \bar U} w_{j,l}$, we obtain 
\(
&       \bigg[\sum_{j\in U} \bigg \{ \frac{\tilde e_j}{d_j} S_t(j) \bigg\}\bigg]   \sum_{j\in U} d_j  
 =     \sum_{j\in U}  \bigg \{ \tilde {e_j} S_t(j) \frac{ \sum_{j'\in U} d_{j'} }{d_j}\bigg \}\\
& \le   \sum_{j\in U}   {\tilde e_j} \max_{j\in U}\bigg \{ S_t(j) \frac{ \sum_{j'\in U} d_{j'}}{d_j}\bigg \} \le   \Big(\sum_{j\in U}   {\tilde e_j} \Big)\max_{j\in  U\cup \bar U}\bigg \{ S_t(j) \frac{ \sum_{j'\in U\cup \bar U} d_{j'}}{d_{j}}\bigg \}.
\)
After rearranging terms, we obtain:
\(
  & \text{Pr}( x_{t+1} \in \bar U\mid x_{\tilde t}\in U \;\forall \tilde{t} \le t) \le   \frac{\sum_{j\in U} \tilde  e_{j}}{ \sum_{j\in U} d_{j}}  \max_{j\in  U\cup \bar U}\bigg \{ S_t(j) \frac{ \sum_{j'} d_{j'} }{d_j}\bigg \}\\
& \le   \frac{\sum_{j\in U}
\tilde e_{j}}{ \min(\sum_{j\in U} d_{j} ,\sum_{j\in \bar U} d_{j} )}  \max_{j\in
 U\cup \bar U}\bigg \{ S_t(j) \frac{
\sum_{j'} d_{j'} }{d_j}\bigg \}.
\)
Using $\pi_j= d_j/\sum_{j'} d_{j'}$ and
taking the minimum over both sides, we have
\(
\min_{U:|U|\le m-1} \text{Pr}( x_{t+1} \in \bar U\mid x_{\tilde t}\in U \;\forall \tilde{t}\le t) \le\sqrt{\lambda_2(\mathcal L)}  
\max_{j} S_t(j) /\pi_j.
\)
Letting $U_{k^*}$ be one that reaches the minimum on the left-hand side, by a similar argument in the proof of Theorem 2, 
\(
\mathbb{E}(\hat t_{k^*+1} - \hat t^*_k) \ge \frac{1}{M\sqrt{\lambda_2(\mathcal
L)}}.
\)
Adding the other $(m-2)$ steps, each with $\hat t_{k'+1} - \hat t_{k'}\ge 1$, leads to the result.
\end{proof}
\subsection*{Proof for Theorem 4}

\begin{proof}
The Neumann series converges if the spectral radius $\lambda_1(P_{U,U}) <1$ strictly. Since $P_{U,U}$ is a non-negative matrix, by the Perron–-Frobenius theorem, $\lambda_1(P_{U,U})\le \max_i \sum_{j} p_{i,j}\le 1$.

Since $P_{U,U}$ is irreducible, by the Perron–-Frobenius theorem, there exists a unique vector $P^{\rm T}_{U,U} \phi_* = \lambda_1(P_{U,U}) \phi_*$, with $\phi_*$ all positive and $1^{\rm T} \phi_*=1$. 
 We follow Chapter 8 of \cite{meyer2023matrix}, and let $Q$ be a non-negative matrix such that $P^{\rm T}_{U,U}+Q$ has each column summable to 1, hence $1^{\rm T} (P^{\rm T}_{U,U} + Q) \phi\le 1$ for any $\phi$ all positive and $1^{\rm T}\phi=1$. Since there exists 
at least $\eta_j>0$, we know at least one $Q_{j,l}>0$.

If $\lambda_1(P_{U,U}) = 1$, we would have
\(
1^{\rm T} (P^{\rm T}_{U,U} + Q) \phi_* = 1^{\rm T}  \phi_* + 1^{\rm T} Q \phi_*>1,
\)
which is a contradiction. Therefore, we know $\lambda_1(P_{U,U}) <1$.
\end{proof}

\bibliographystyle{biometrika}
\bibliography{ref}
\appendixtwo
\section*{Supplementary Materials}

\subsection{Runtime comparisons via number of random walk steps}

\begin{figure}[H]
    \centering
    \begin{subfigure}[t]{0.32\textwidth}
        \centering
        \includegraphics[width=1\textwidth]{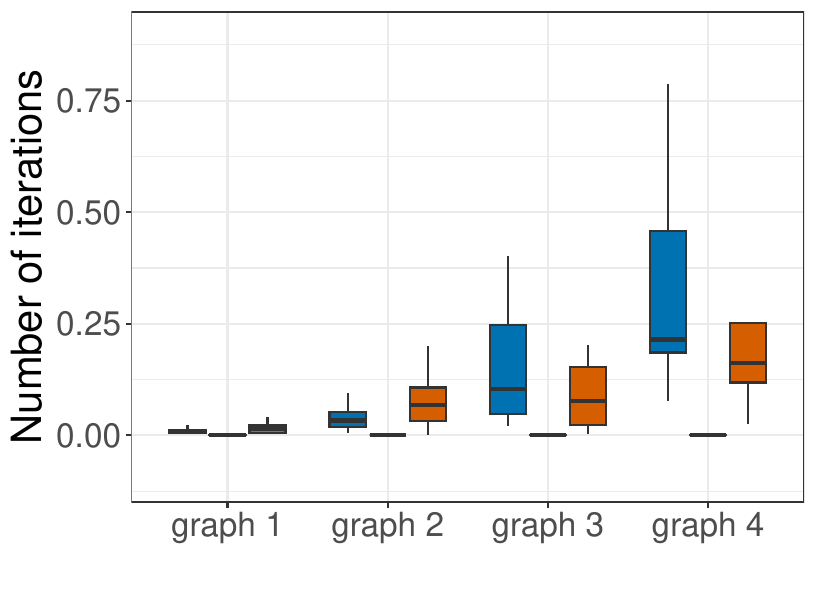}
        \caption{Number of iterations to finish over different
        bottleneck rates $1/\sqrt{\lambda_2(\mathcal L)}$.}
    \end{subfigure}\;
    \begin{subfigure}[t]{0.32\textwidth}
        \centering
        \includegraphics[width=1\textwidth]{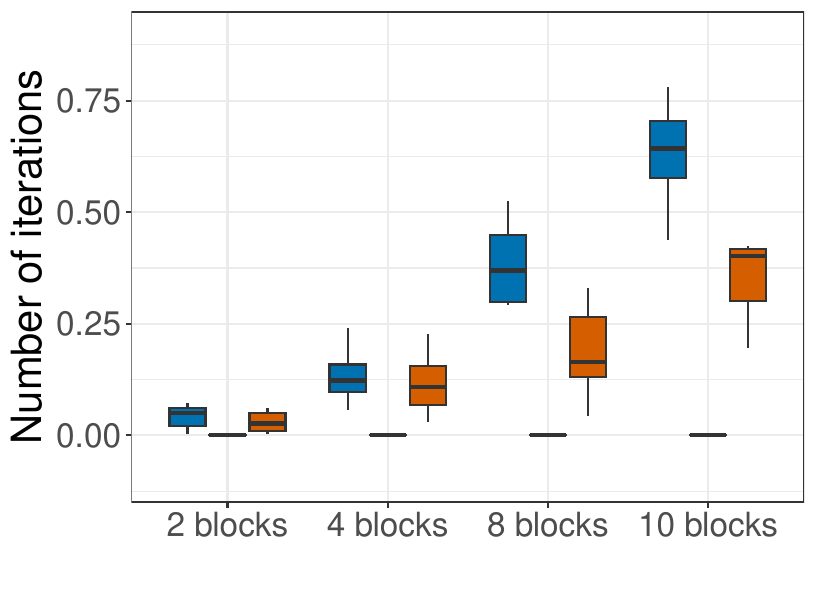}
        \caption{Number of iterations to finish over different numbers of blocks (hence the numbers of bottlenecks).}
    \end{subfigure}\;
    \begin{subfigure}[t]{0.32\textwidth}
        \centering
        \includegraphics[width=1\textwidth]{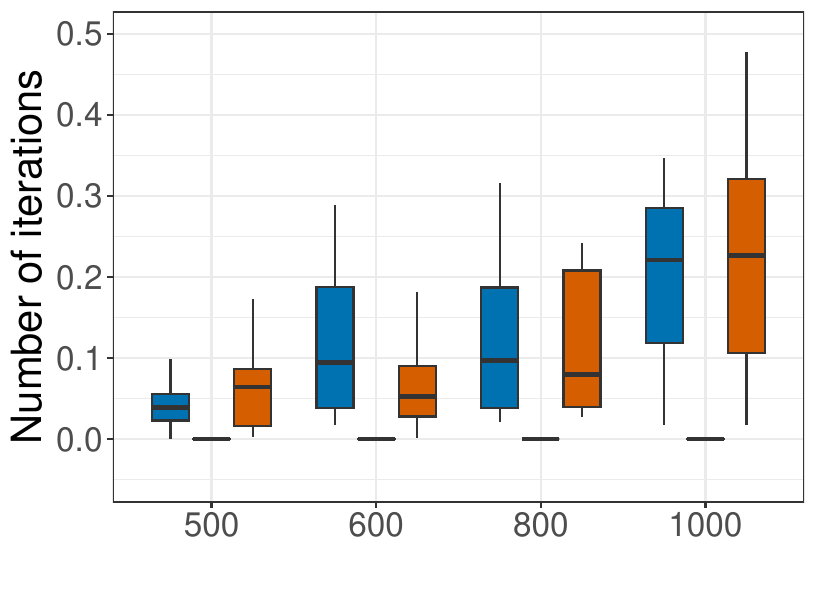}
        \caption{Number of iterations to finish over different numbers of nodes.}
    \end{subfigure}
    \caption{Comparisons of number of iterations to finish between random walk cover (blue) and fast-forwarded cover (grey) algorithm.}
    \label{fig:iterations}
\end{figure}

\subsection{Spanning tree-augmented dendrogram Gibbs sampler}

The proposed Gibbs sampler iterates between the following sampling steps: 
\begin{itemize}
	\item  Update the undirected version of the spanning tree $T$ from the full conditional $\Pi(T \mid - )$, which is a spanning tree distribution of product-over-edges form, using the fast-forwarded cover algorithm. Form a directed tree $\vec T$ from $T$ by assigning $0$ as the root, and having all edges pointed away from $0$;
	\item  For $i=1,\ldots,n$,  update $z_i$ from the full conditional $\Pi(z_i=k \mid -)$, a categorical distribution over $1, \cdots \tilde m$;
	\item 
    Update $(\pi_1, \cdots, \pi_{\tilde m})$ from the full conditional $\Pi(\pi_1, \cdots, \pi_{\tilde m} \mid - )$, a Dirichlet distribution;
     \item 
    Update the $\mu_k$'s and $\Sigma$ from the full conditional $\Pi(\mu_2, \cdots, \mu_{\tilde m}, \Sigma \mid - )$, which takes the form of a normal-inverse-Wishart distribution. 
\end{itemize}
Recall that the root parameter $\mu_1$ is set at $0$. Once tree samples have been collected, pruning steps outlined in section \ref{sec:dendrogram} are performed to obtain the reduced dendrograms.

\subsection{Details on plots on posterior similarity matrices and Massachusetts map}

Posterior similarity matrices at depths $1, 2, 3$ were generated for $n = 123$ Massachusetts communities available in the crimes and communities dataset. Each row/column in the matrix corresponds to a community. The $ij$th entry of the posterior similarity matrix at depth $R$ corresponds to the fraction of Gibbs sampler iterations for which the two communities share an ancestor node at depth $R$. The posterior similarity matrix was computed on samples $3501$ to $5000$ with thinning factor $10$. The depth $1$ posterior similarity matrix is then reordered according to the output of a spectral biclustering algorithm to reveal block-diagonal structures. The same ordering is applied to all 3 posterior similarity matrices at depths $1, 2, 3$.  Maps data were obtained from the Massachusetts government website. The data from the crimes and communities dataset do not include all Massachusetts towns/cities. Counties that are missing from the dataset are set to transparent on the map. 

\subsection{Details for reversible-jump Markov chain Monte Carlo}
We use $V_{\vec T}$ as the set of nodes in the dendrogram, and $V_{\vec T}^C= (1,\ldots, \tilde m) \setminus V_{\vec T}$ as the other nodes not on the dendrogram. Further, we use $V^{\text{empty-leaf}}_{\vec T}$ to denote the node set $\{ k\in V_{\vec T}: k \text{ has no children}, n_k=0\}$.

We use the same full conditional distributions as in the main text to update $(\pi_1,\ldots, \pi_{\tilde m})$, and those $\mu_k$ with $k\in V_{\vec T}$. When updating $z_i$, we use the same multinomial distribution, except with $\phi(y_i; \mu_k, \Sigma)$ replaced by zero density if $k\in V_{\vec T}^C$.

 For updating the dendrogram, we use the proposals of birth and death steps in each iteration. We draw a binary $u\sim \text{Bern}(p_0)$: if $u=1$, we use the birth proposal unless $|V_{\vec T}|=n$; if $u=0$, we use the death proposal unless $|V_{\vec T}|=1$. Therefore, we have the birth probability $p_{\text{birth}}(\vec T) = p_0$ if $|V_{\vec T}|<n$, and $0$ otherwise; and the death probability $p_{\text{death}}(\vec T) = 1- p_0$ if $|V_{\vec T}|>  1$, and $0$ otherwise. In our simulations, we set $p_0 = 0.1$. 
\begin{itemize}
	\item (Birth)  Grow a new edge: draw $j$ uniformly from $V_{\vec T}$, find the smallest index $l$ from $V^C_{\vec T}
    $, draw $\mu_l \sim \mathcal G(\mu_l\mid \mu_j)$, and attach $(j\to l)$ to $\vec T$ to form a proposed $\vec T^*$. Accept the $\vec T^*$ with probability
	\[ \min \bigg \{1,  \frac{0.01  p_{\text{death}}(\vec T^*)  | V^{\text{empty-leaf}}_{\vec T^*}|^{-1} }{p_{\text{birth}}(\vec T) |V_{\vec T}|^{-1} }\bigg \}.\]
	\item (Death)  Prune an empty leaf node: draw $j$ uniformly from $ V^{\text{empty-leaf}}_{\vec T}$, remove $j$ from $\vec T$ to form a proposed $T^*$. Accept the $\vec T^*$ with probability
	\[ \min \bigg \{1,  \frac{p_{\text{birth}}(\vec T^*)  |V_{\vec T^*}|^{-1}  }{0.01 p_{\text{death}}(\vec T) | V^{\text{empty-leaf}}_{\vec T}|^{-1} }\bigg \},\]
	provided that the current $\vec T$ has $|V^{\text{empty-leaf}}_{\vec T}| >0$. If $|V^{\text{empty-leaf}}_{\vec T}| =0$ skip the step.
\end{itemize}
In the above, the Metropolis-Hastings acceptance ratio has a simple form because in the birth step, the product of the likelihood  at ${\vec T}$ and the proposal density for $\mu_l$  satisfy $L(y; (\mu_j)_{j \in V_{\vec T}}, \vec T, \cdot) \mathcal G(\mu_l\mid \mu_j)= L(y; (\mu_j)_{j \in V_{\vec T^*}}, \vec T^*, \cdot)$, hence canceling out the likelihood at the proposed state; and the transform $\{ (\mu_j)_{j \in V_{\vec T}}, \mu_l \} \leftrightarrow (\mu_j)_{j \in V_{\vec T^*}}$ is one-to-one and has Jacobian determinant 1.

\subsection{Further results on the Gibbs Sampler}

\begin{figure}[!htb]
\begin{subfigure}[t]{.32\textwidth}
   \begin{overpic}[width=\textwidth]{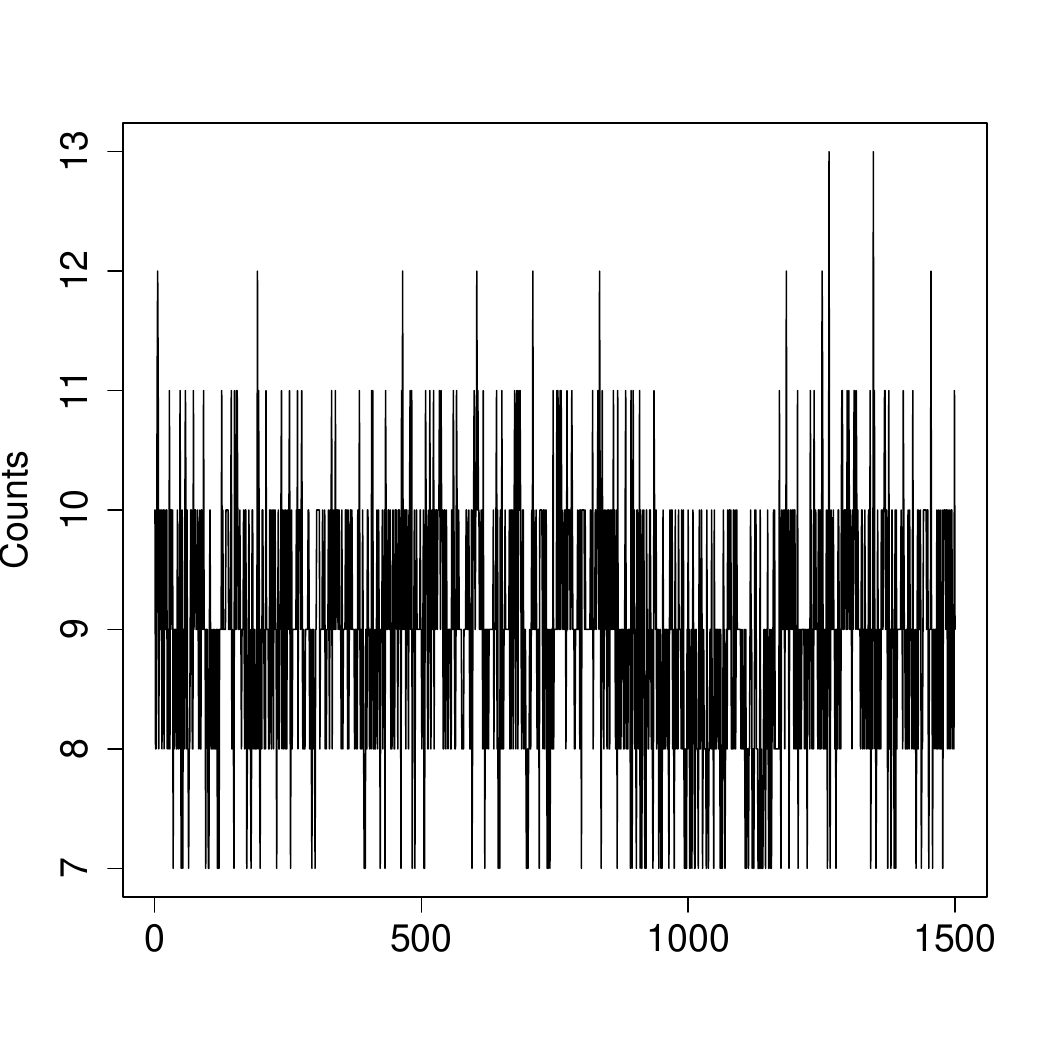}
   \put(45, -1){\scriptsize Iteration}
   \end{overpic}
   \caption{\scriptsize Trace plot of depth.}
\end{subfigure}
\begin{subfigure}[t]{.32\textwidth}
   \begin{overpic}[width=\textwidth]{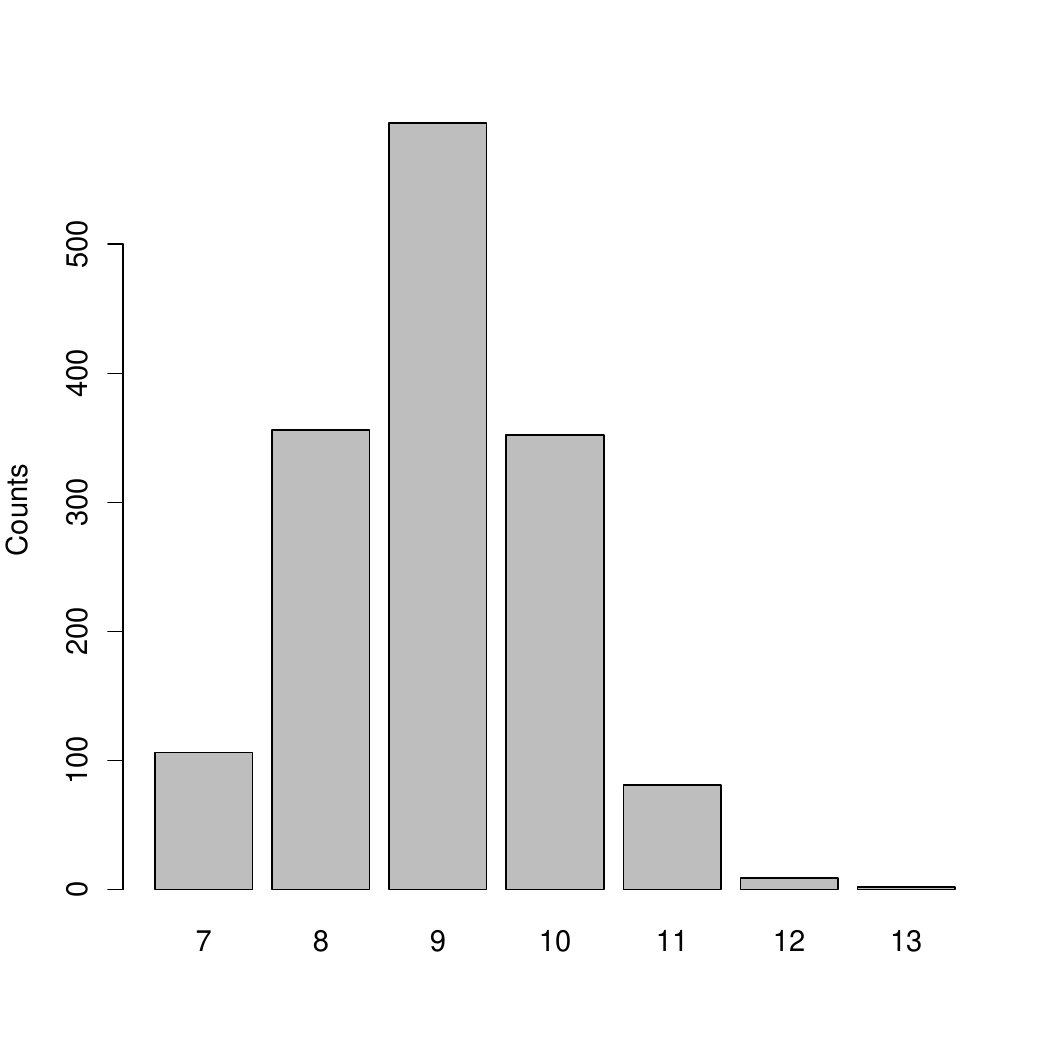}
   \put(50, -1){\scriptsize Depth}
   \end{overpic}
   \caption{\scriptsize Bar plot of depth.}
\end{subfigure}
\begin{subfigure}[t]{.32\textwidth}
   \begin{overpic}[width=\textwidth]{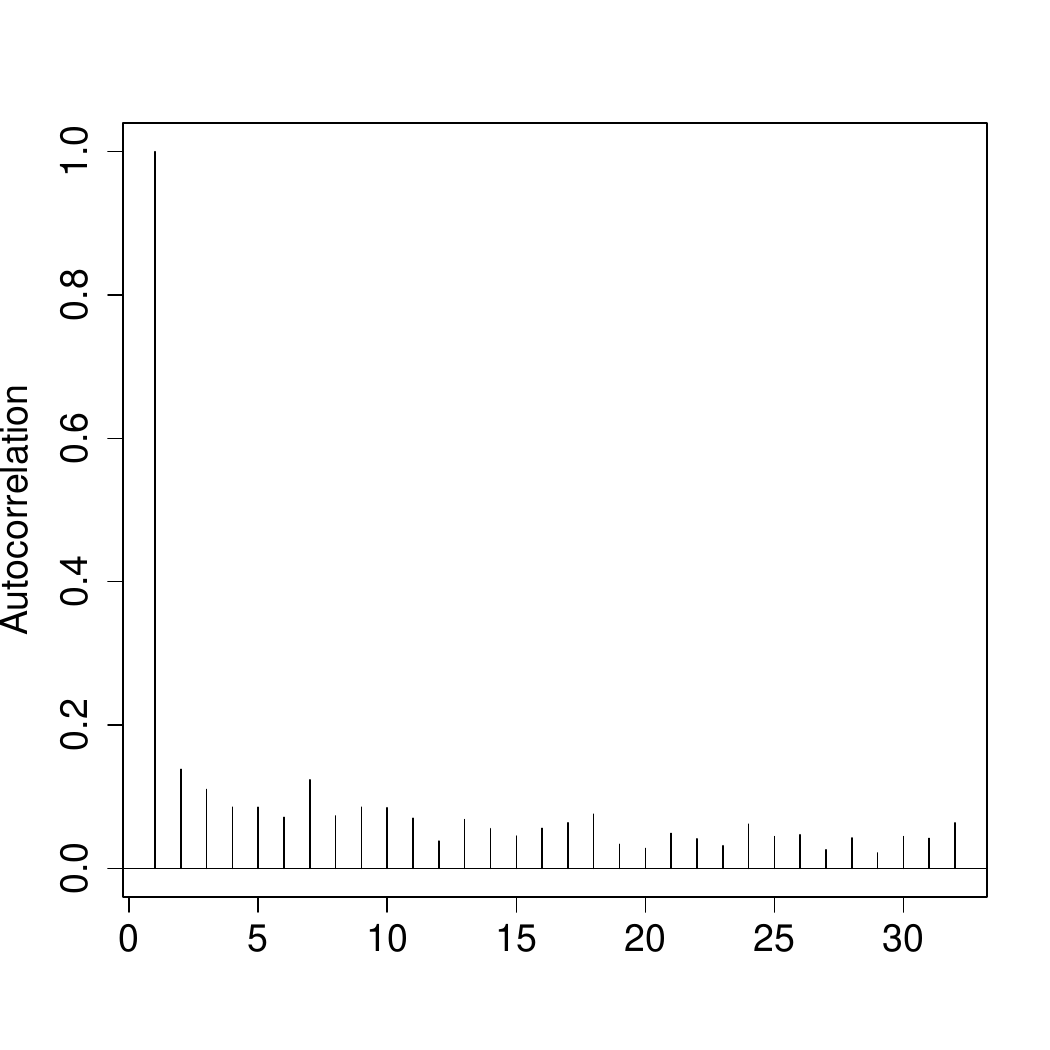}
   \put(50, -1){\scriptsize Lag}
   \end{overpic}
   \caption{\scriptsize Autocorrelation plot of depth.}
\end{subfigure}
\caption{Trace, bar and autocorrelation plots for the depth of the tree samples from the Gibbs sampler.}
\end{figure}

\begin{figure}[!htb]
\begin{subfigure}[t]{.32\textwidth}
   \begin{overpic}[width=\textwidth]{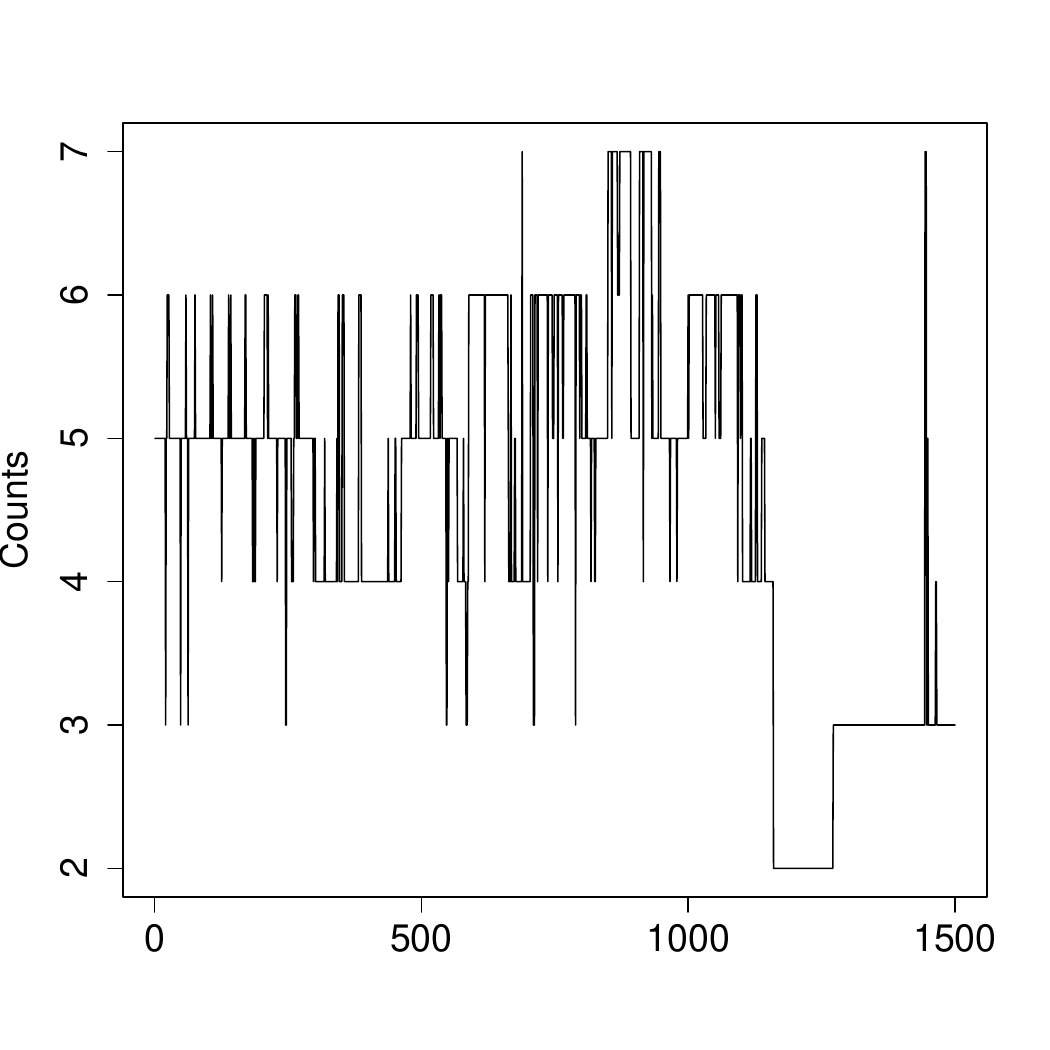}
   \put(45, -1){\scriptsize Iteration}
   \end{overpic}
   \caption{\scriptsize Trace plot of depth.}
\end{subfigure}
\begin{subfigure}[t]{.32\textwidth}
   \begin{overpic}[width=\textwidth]{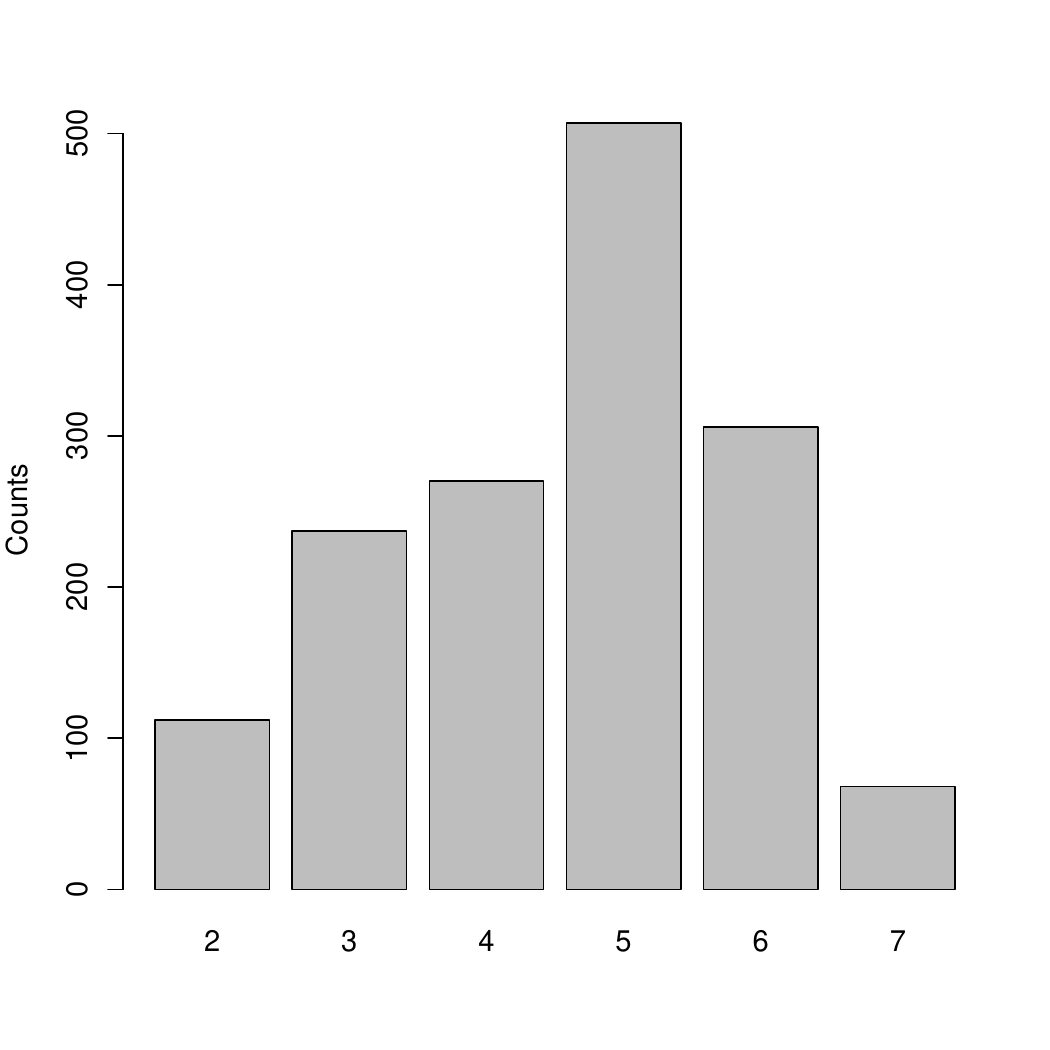}
   \put(50, -1){\scriptsize Depth}
   \end{overpic}
   \caption{\scriptsize Bar plot of depth.}
\end{subfigure}
\begin{subfigure}[t]{.32\textwidth}
   \begin{overpic}[width=\textwidth]{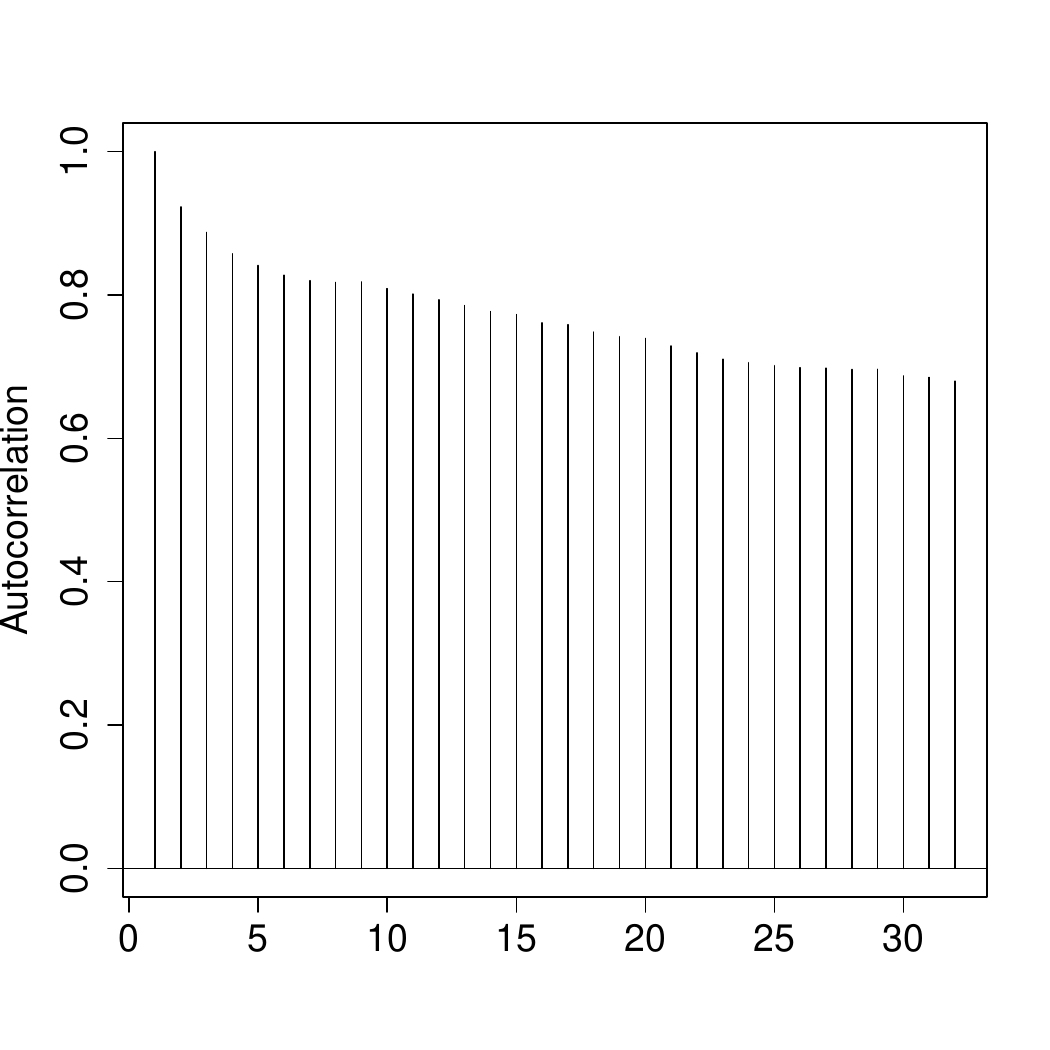}
       \put(50, -1){\scriptsize Lag}
   \end{overpic}
   \caption{\scriptsize Autocorrelation plot of depth.}
\end{subfigure}

\caption{Trace, bar and autocorrelation plots for the depth of the tree samples from the reversible-jump Markov chain Monte Carlo sampler.}
\end{figure}

\subsection{Comparisons with Laplacian-based algorithm}
\begin{figure}[H]
    \centering
    \begin{subfigure}[t]{0.46\textwidth}
        \centering
        \includegraphics[width=1\textwidth]{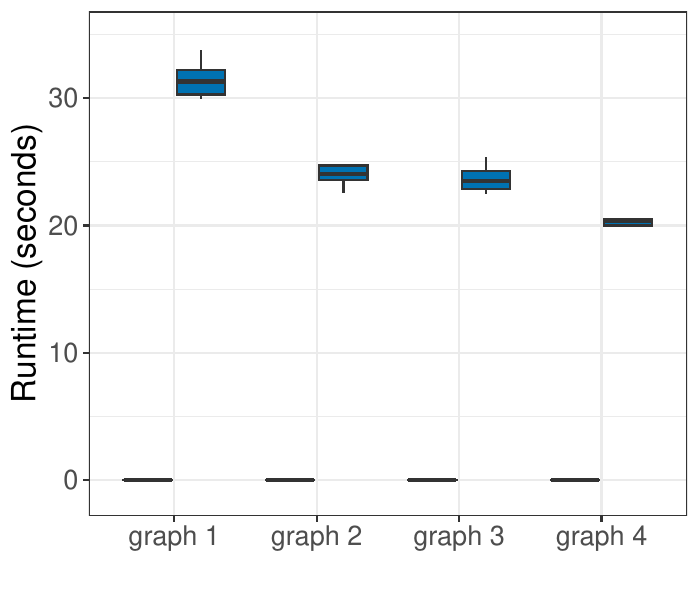}
        \caption{Runtime for different bottleneck size $1/\sqrt{\lambda_2(\mathcal L)}$.}
    \end{subfigure}\;
    \begin{subfigure}[t]{0.46\textwidth}
       
        \centering
        \includegraphics[width=1\textwidth]{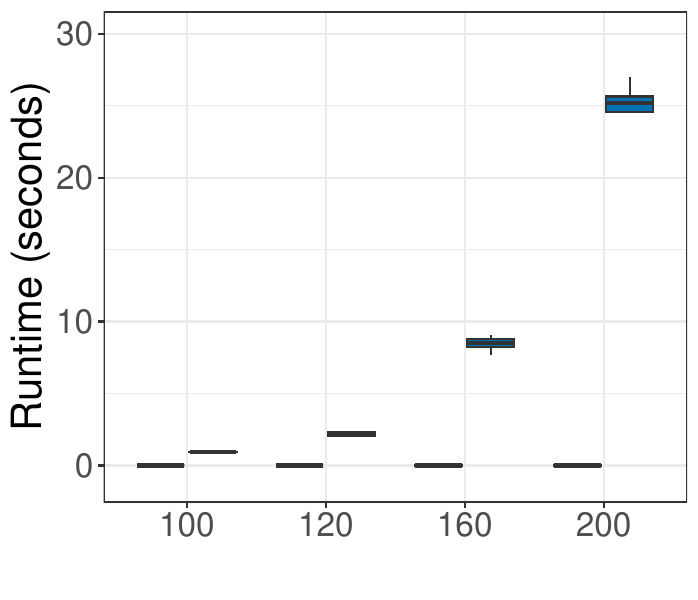}
        \caption{Runtime over different numbers of nodes.}
    \end{subfigure}
    \caption{Comparisons of runtimes between the Laplacian-based (blue) and the fast-forwarded cover (grey) algorithms. Some boxes appear close to a thin line due to the small variance in running time in the simulation.}     \label{fig:laplace_runtime}
\end{figure}

We simulate a graph with $m = 100$ nodes, where we partition the nodes $V$ into two blocks with $|V_1|=|V_2|=50$. The graph is represented by a symmetric weight matrix $W\in\mathbb{R}^{100\times 100}$ with each $w_{j,l}=b_{j,l}$.  We set $b_{j,l}=1$ for those $j$ and $l$ that are in the same node partition, and $b_{j,l} \sim \text{Bern}(\zeta)$ for those $j$ and $l$ that are in different partitions. We conducted experiments at different edge densities with $\zeta$ from $(0.5,0.1,0.05,0.01)$, corresponding to graphs 1 to 4 in Figure \ref{fig:laplace_runtime}(a). Under each value of $\zeta$, we draw 10 spanning trees for both the fast-forwarded cover algorithm and a Laplacian-based algorithm outlined in Algorithm 1 of \cite{harvey2016Generating}, which has roots in \cite{guenoche1983random, kulkarni1990Generating}. Algorithm 1 as stated in \cite{harvey2016Generating} generates uniform spanning tress from unweighted graphs. 
Other theoretical algorithms which use Laplacian solvers, including \cite{madry2014fast,schild2018almost}, generally target approximate random spanning tree sampling or impose assumptions on edge weights.

Figure \ref{fig:laplace_runtime}(a) shows that the runtimes for both algorithms do not change substantially with bottleneck size, although the fast-forwarded cover algorithm is much faster. 
We then run the fast-forwarded cover and Laplacian-based algorithms on graphs with numbers of nodes ranging within $(100,120,160,200)$ in the two block case using $b_{j,l}=b_{l,j}\sim \text{Bern}(0.1)$. Figure \ref{fig:laplace_runtime}(b) shows that the Laplacian-based algorithm slows down much more rapidly than the fast-forwarded cover algorithm.

\subsection{Posterior sampling via subtree prune and regraft}

We adopt a setting similar to the reversible jump sampler, where the only difference is in the sampling of the tree component of the model. 
For sampling the tree component $\vec{T}$, we use a subtree prune and regraft move, which we define by the following two steps. 
\begin{itemize}
	\item (Pruning)  Draw a non-root pruning node $\omega$ uniformly  from the node set of the current spanning tree $\vec{T}$. Remove the edge connecting $\omega$ to its parent. This splits $\vec{T}$ into two trees. The induced subgraph formed by $\omega$ and its descendants is denoted $\vec{T}_1$, and the induced subgraph formed by the remaining nodes in $\vec{T}$ is denoted $\vec{T}_2$. 
	\item (Regraft)  Draw a regraft node $\zeta$ uniformly at random from the node set of $\vec{T}_2$. Join $\vec{T}_2$ and $\vec{T}_1$ by adding the directed edge $(\zeta \to \omega)$. This forms the proposed tree $\vec{T'}$.
\end{itemize}

The proposal acceptance rate has the form 
$$\min\bigg\{1, \frac{L(y; (\mu_j)_{j \in V_{\vec T^*}}, \vec T^*, \cdot)}{L(y; (\mu_j)_{j \in V_{\vec T}}, \vec T, \cdot)}\bigg\}.$$

\begin{figure}[H]
\begin{subfigure}[t]{.4\textwidth}
    \begin{overpic}[width=\textwidth]{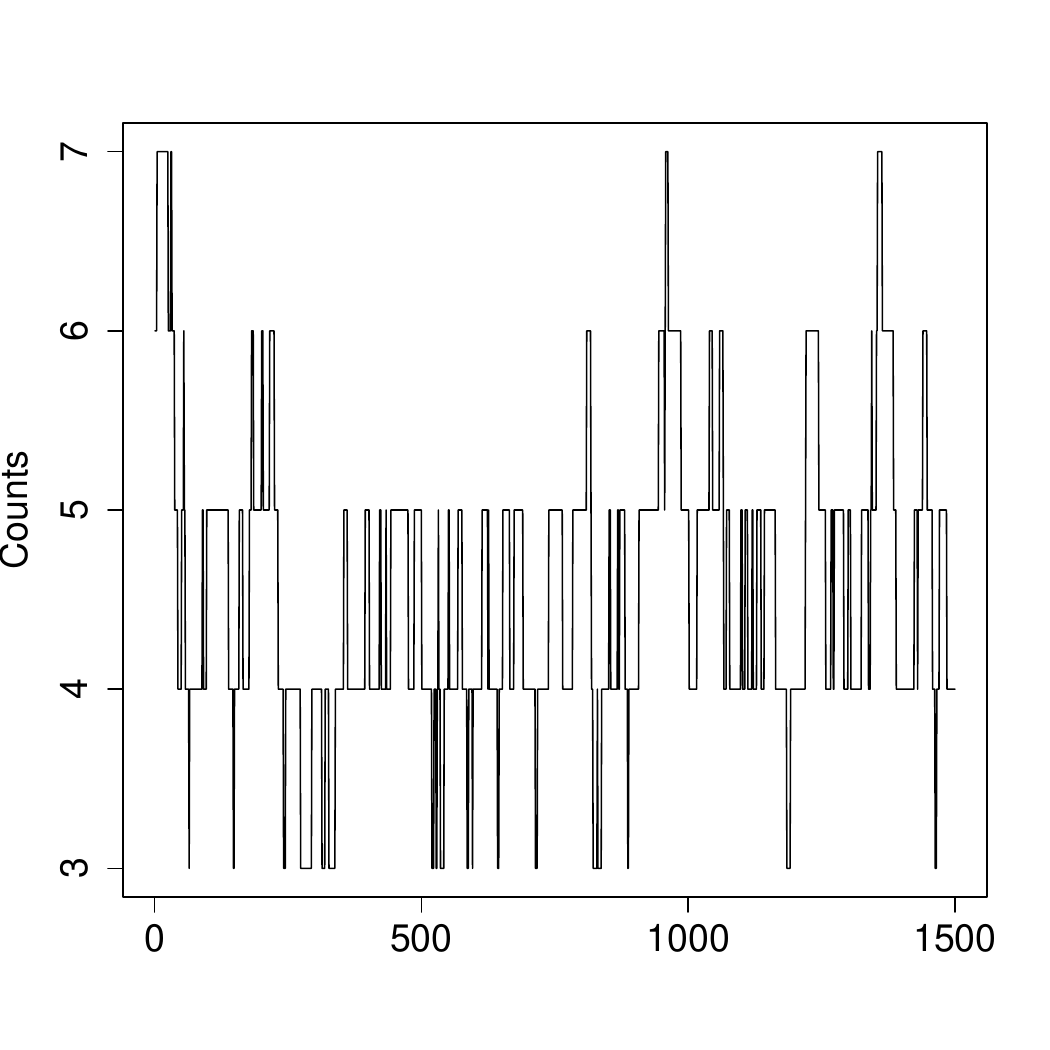}
    \put(45, -1){\scriptsize Iteration}
    \end{overpic}
    \caption{Trace of maximum degree from subtree-prune-regraft sampler.}
\end{subfigure}\;
\begin{subfigure}[t]{.4\textwidth}
    \begin{overpic}[width=\textwidth]{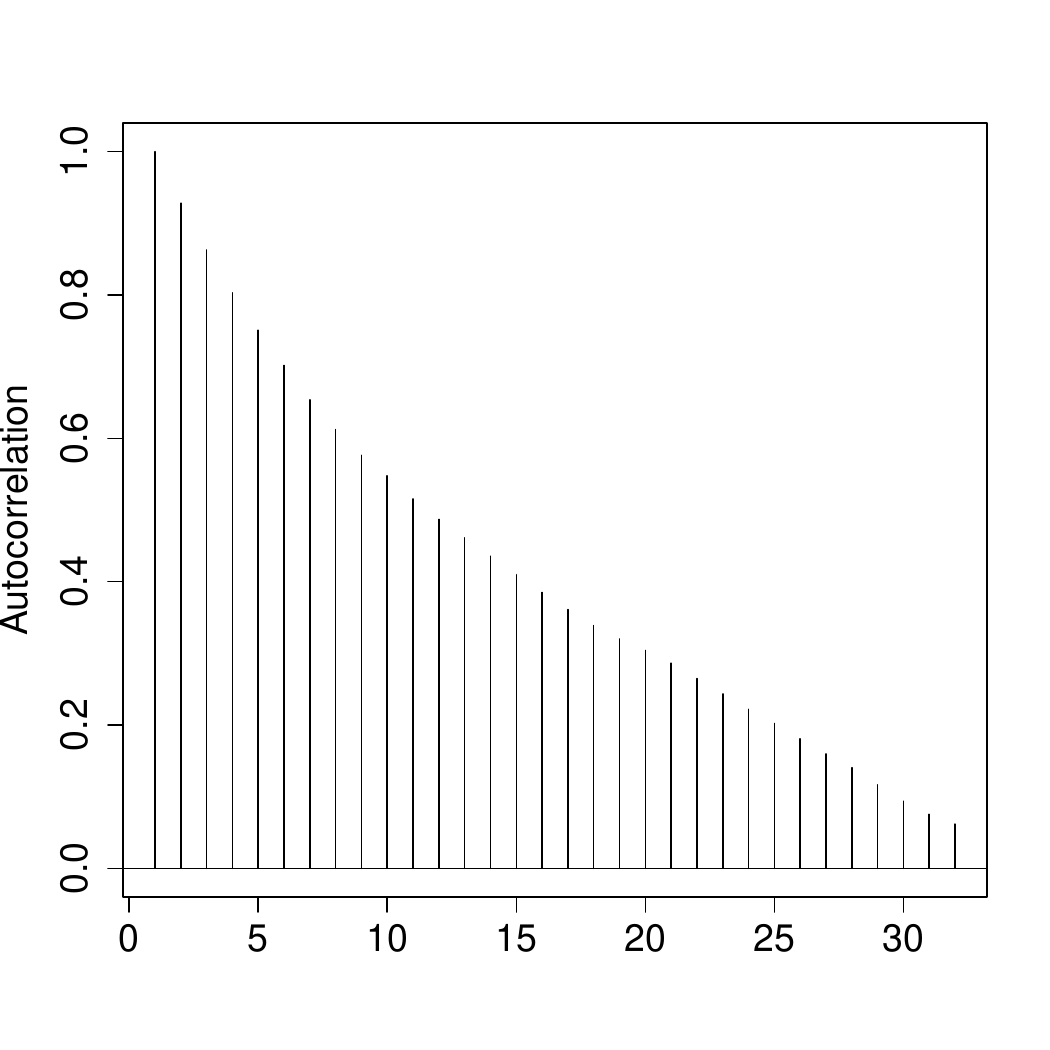}
    \put(50, -1){\scriptsize Lag}
    \end{overpic}
    \caption{Autocorrelation of maximum degree from subtree-prune-regraft sampler.}
\end{subfigure}\\
\begin{subfigure}[t]{.4\textwidth}
    \begin{overpic}[width=\textwidth]{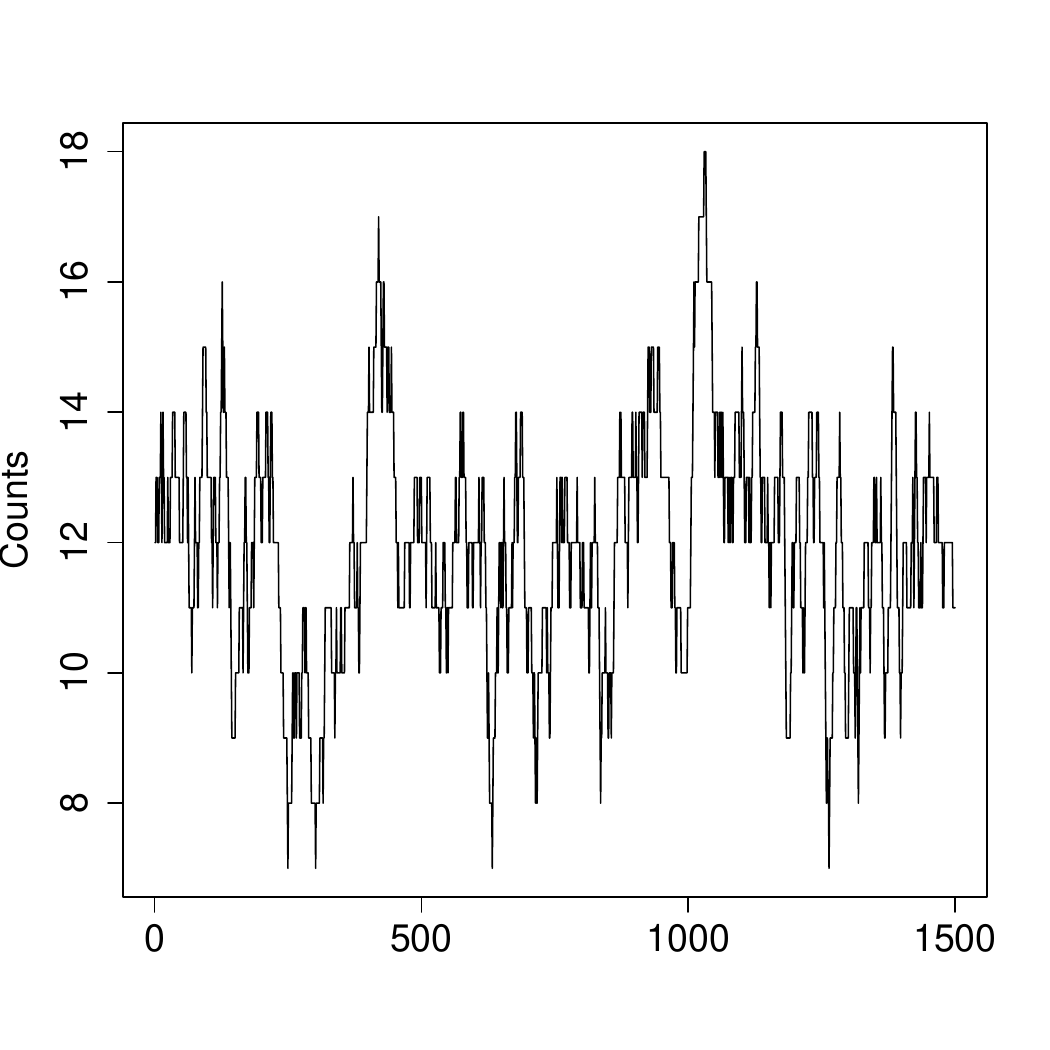}
    \put(45, -1){\scriptsize Iteration}
    \end{overpic}
    \caption{Trace of number of leaves from subtree-prune-regraft sampler.}
\end{subfigure}\;
\begin{subfigure}[t]{.4\textwidth}
    \begin{overpic}[width=\textwidth]{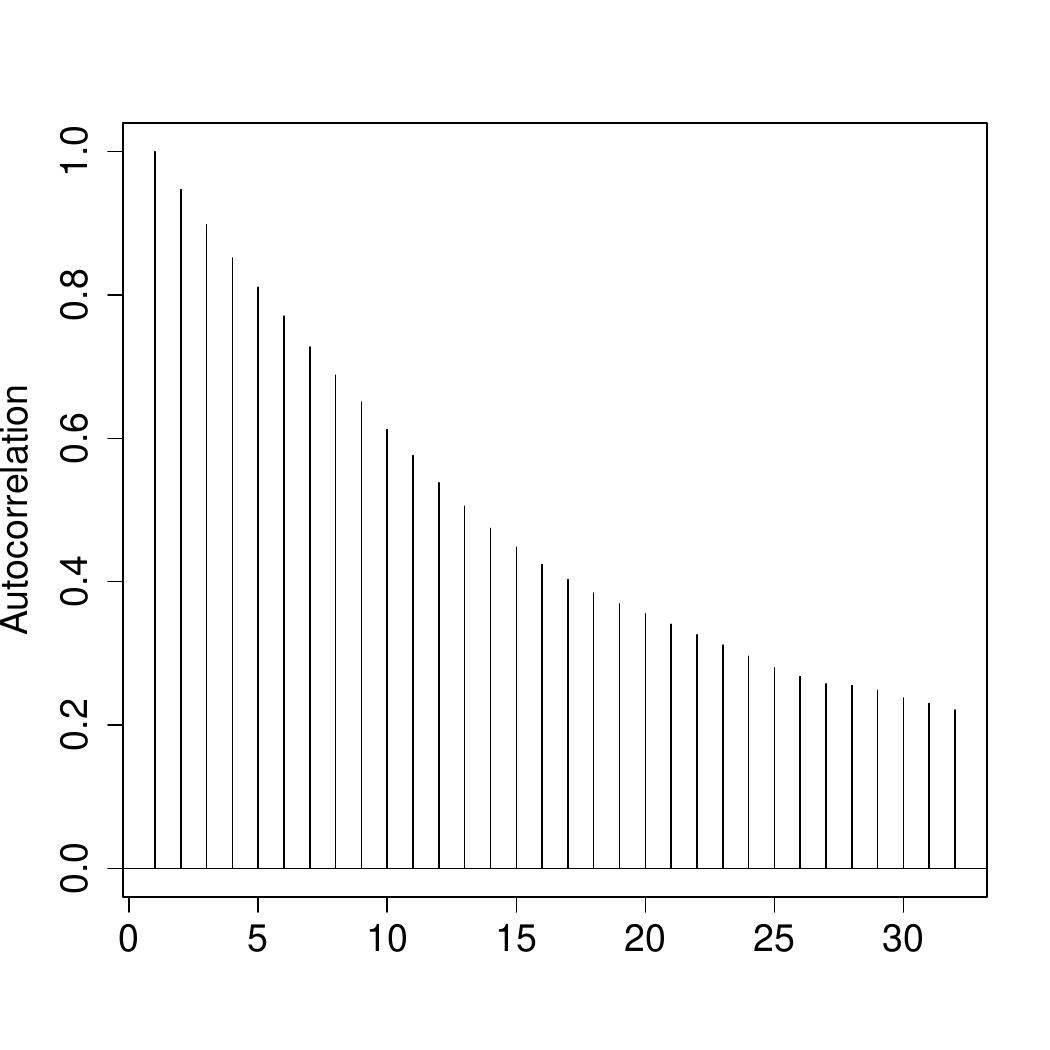}
    \put(50, -1){\scriptsize Lag}
    \end{overpic}
    \caption{Autocorrelation of number of leaves from subtree-prune-regraft sampler. }
\end{subfigure}\\
\begin{subfigure}[t]{.4\textwidth}
    \begin{overpic}[width=\textwidth]{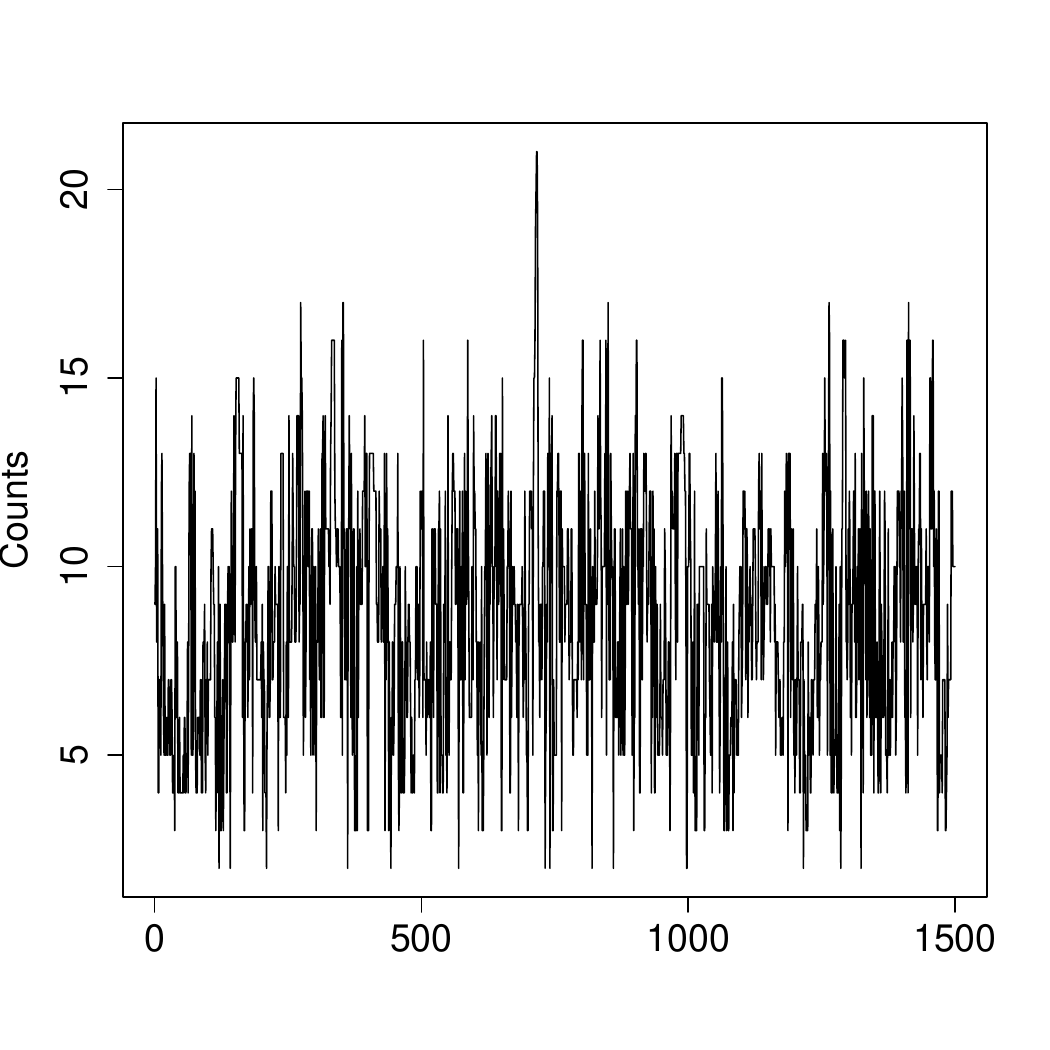}
    \put(45, -1){\scriptsize Iteration}
    \end{overpic}
    \caption{Trace of depth from subtree-prune-regraft sampler.}
\end{subfigure}\;
\begin{subfigure}[t]{.4\textwidth}
    \begin{overpic}[width=\textwidth]{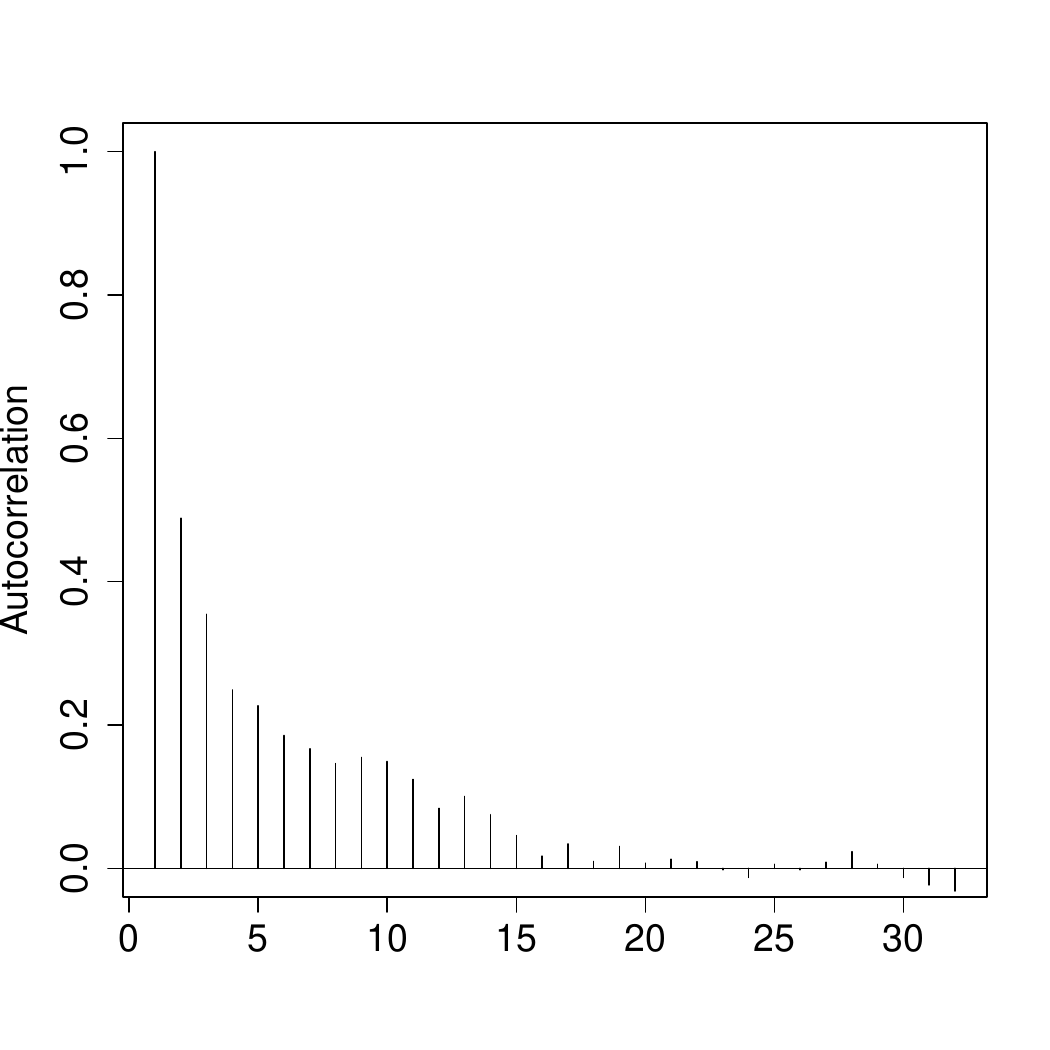}
    \put(50, -1){\scriptsize Lag}
    \end{overpic}
    \caption{Autocorrelation of depth from subtree-prune-regraft sampler.}
\end{subfigure}
\caption{The subtree-prune-regraft sampler for a spanning tree-augmented dendrogram model.\label{fig:spr}}
\end{figure}

Figure \ref{fig:spr} shows the mixing performance. 
The per sample effective sample sizes for maximum degree, maximum depth and number of leaves under the subtree-prune-regraft sampler are $0.037, 0.200, 0.027$ respectively. The mixing is slower than that of the Gibbs sampler.

\subsection{Normalized Laplacian}

For any non-negative weight matrix $W$, define the normalized Laplacian \citep{chung1997spectral} as 
\(
\mathcal{L}=I-\frac{1}{2}\big\{\Phi^{1 / 2} (D^{-1}W) \Phi^{-1 / 2}+\Phi^{-1 / 2} (W^{\rm T}D^{-1}) \Phi^{1 / 2}\big\},
\)
where $\Phi =\text{diag}(\pi_j)_{j=1}^m$ and $D=\text{diag}(d_j)_{j=1}^m$. Since $\sum_{l=1}^m \pi_j(w_{j,l} /d_j)= \sum_{k=1}^m \pi_k(w_{k,j} /d_k)$ holds for
%\[\label{eq:inv_circulation}
$\pi_j= d_j/ (\sum_{k=1}^m d_{k}),$
%\] 
the normalized Laplacian reduces to
\(
\mathcal{L}=I-D^{-1/2}\frac{W+W^{\rm T}}{2}D^{-1/2}.
\)
The above form further reduces to $\mathcal{L}=I-D^{-1/2}WD^{-1/2}$ if $W=W^{\rm T}$.

\end{document}